\documentclass[12pt]{book}
\usepackage[height=8.85in,width=6.45in]{geometry}

\usepackage{times}

\pdfoutput=1

\usepackage[utf8]{inputenc}
\usepackage[svgnames]{xcolor}
\usepackage{graphicx}
\usepackage{appendix}

\usepackage{ascmac}
\usepackage{amsmath}
\usepackage{amssymb}
\usepackage{mathtools}
\usepackage{bm}

\usepackage{braket}
\usepackage{cancel}
\usepackage{fancybox}
\usepackage{slashed}

\usepackage{multirow}
\usepackage{tikz}
\usepackage{tikz-cd}
\usetikzlibrary{shadows.blur}
\usetikzlibrary{fit}
\usetikzlibrary{shapes.misc}
\usetikzlibrary{decorations.pathmorphing}

\usepackage{slashed}
\usepackage{array}
\usepackage{extarrows}
\usepackage{colortbl}
\usepackage{cite}
\usepackage
[
	bookmarks=false,
	colorlinks=true,
	citecolor=refcolor,
	linkcolor=refcolor
]{hyperref}
\usepackage{subfig}
\usetikzlibrary{shapes.geometric}
\usetikzlibrary{positioning}
\usepackage[sectionbib]{chapterbib}
\usepackage{floatrow}
\DeclareColorBox{gray}{\colorbox{black!7}}
\usepackage{dashbox}
\usepackage{CJKutf8}

\definecolor{Blue}{rgb}{0,0,1}
\definecolor{blue}{rgb}{0.0, 0.4, 1}
\definecolor{darkgreen}{rgb}{0.,0.6,0.}
\definecolor{lightyellow}{rgb}{1.0, 0.95, 0.7}
\definecolor{lightblue}{rgb}{0.7, 0.9, 1.0}
\definecolor{lightpink}{rgb}{1.0, 0.85, 0.95}
\definecolor{lightgreen}{rgb}{0.7, 1.0, 0.4}
\definecolor{refcolor}{rgb}{0.3,0.3,1}

\newcommand*{\magenta}[1]{\textcolor{magenta}{#1}}

\definecolor{colorF}{rgb}{0,0.72,0.92}
\definecolor{colorG}{rgb}{0,0,1}
\definecolor{colorH}{rgb}{0.72, 0, 0.92}
\definecolor{colorI}{rgb}{0.92, 0.2, 0.8}

\definecolor{colorA}{rgb}{0,0,1}
\definecolor{colorB}{rgb}{1,0.5,0}
\definecolor{colorC}{rgb}{0,0.72,0.92}
\definecolor{colorD}{rgb}{1,0.5,0}
\definecolor{colorE}{rgb}{0,0.72,0.92}

\newcommand*{\bZ}{\mathbb{Z}}
\newcommand*{\bR}{\mathbb{R}}
\newcommand*{\bC}{\mathbb{C}}

\newcommand*{\cA}{\mathcal{A}}
\newcommand*{\cB}{\mathcal{B}}

\newcommand*{\cE}{\mathcal{E}}

\newcommand*{\cG}{\mathcal{G}}

\newcommand*{\cI}{\mathcal{I}}

\newcommand*{\cK}{\mathcal{K}}
\newcommand*{\cL}{\mathcal{L}}

\newcommand*{\cN}{\mathcal{N}}
\newcommand*{\cO}{\mathcal{O}}

\newcommand*{\cQ}{\mathcal{Q}}

\newcommand*{\vev}[1]{\langle #1 \rangle}

\usepackage{spectralsequences}
\DeclareSseqGroup\tower {} {
	\class(0,0)\foreach \i in {1,...,6} {
		\class(0,\i)
		\structline(0,\i-1,-1)(0,\i,-1)
	}
}
\usepackage{arydshln}

\usepackage{fancyhdr}
\usepackage[multiple]{footmisc}

\begin{document}

\begin{titlepage}

	\begin{center}
		Doctoral Dissertation\vspace{2mm}
	\end{center}
	
	\bigskip
	
	\begin{center}
	{\Large {\bfseries
		Majorana Fermion Zero Modes\vspace{3mm}\\
		and Anomalies in String and M Theories}
	}\\
 
  \vskip 1.2cm
  
 Yotaro Sato

 \vskip 0.4cm
  
  {\slshape
    Kavli Institute for the Physics and Mathematics of the Universe (WPI), \\
    the University of Tokyo, Kashiwa-no-ha 5-1-5, 277-8583, Japan
   }

 \end{center}
 \vskip 2.5cm
  
\noindent 
In this thesis, we propose a new consistency condition of quantum field theory: Having an odd number of Majorana fermion zero modes on a dynamical point-like soliton signifies an inconsistency in a theory with $3+1$ or higher dimensions.
This inconsistency is related to the non-perturbative anomaly in known models of quantum field theory.

As an application, we check the consistency of string and M theory compactifications with branes. We propose a new criterion on the consistency of brane configurations: An odd number of Majorana fermion zero modes on a D-brane stretching infinitely in space is acceptable, whereas compactifying it to a point-like object leads to an inconsistency. 

We show that a single M5-brane always has an even number of Majorana fermion zero modes because of the symmetry on the worldvolume in $2+1$ or higher dimensional compactifications of M-theory.
We also construct models with an odd number of Majorana fermion zero modes by intersecting D-branes. From our criterion, compactifying such branes to point-like objects is not allowed.
Finally, we show that the gauge invariance on the intersecting branes always requires an even number of Majorana fermion zero modes on a point-like brane. This observation gives another interpretation of the inconsistency we introduced.

 \end{titlepage}

\setcounter{tocdepth}{2}
\tableofcontents
\thispagestyle{empty}

\addtocontents{toc}{\protect\thispagestyle{empty}}


\chapter{Introduction}\label{chap:intro}

\setcounter{page}{1}



\emph{Anomaly} has been a strong tool to analyze quantum field theories and string theories.
One of the key properties of the anomaly is that it can be detected as an infrared effect. The form of anomaly is described by the topology of the theory and does not depend on the details such as the UV structure of the theory. This is called the 't Hooft matching condition \cite{tHooft:1979rat}. This property enables us to investigate new models without knowing the details of them. 


Anomaly is traditionally represented by the violation of symmetries via quantization.
Anomalies of global\footnote{
	The word \emph{global} has various meanings in physics. In this thesis, it means that the symmetry is not gauged.
} symmetries are applied to understand the real physics such as pions and baryogenesis. 
On the other hand, \emph{anomalies of gauge symmetries} are the sign of the inconsistency of the theory. Anomalies of all gauge symmetries must vanish in a consistent quantum field theory. They are applied to speculate the possible form of the beyond Standard Model or the possible vacua of string and M theories.

In general, \emph{anomalies of discrete symmetries} are much harder to analyze than those of continuous symmetries. The abstract classification of anomalies has been developed recently, but there are not many specific constructions of anomalies of discrete parts of gauge symmetries, for example \cite{Witten:1982fp,Wang:2018qoy}. One of them is called the \emph{Witten anomaly}, and is associated with the topology of the gauge group $\pi_4(SU(2))=\bZ_2$. Absence of the Witten anomaly in the heterotic string has been a non-trivial problem \cite{Enoki:2020wel,Tachikawa:2021mby,Yonekura:2022reu}. Since it appears in $SU(2)$ gauge theories, the cancellation of the Witten anomaly is also a non-trivial issue in the Standard Model\footnote{
	The determination of discrete parts of the gauge symmetry in the Standard Model is a subtle problem. See \cite{Tong:2017oea,Davighi:2019rcd,Choi:2023pdp,Reece:2023iqn} for example.
}.
The relation between the Witten anomaly and the fermion zero modes on an 't Hooft-Polyakov monopole was pointed out and discussed in detail in \cite{McGreevy:2011if}. This observation urges us to exploit the fermion zero modes on solitons to understand the anomalies in quantum field theory.

\bigskip

\emph{Fermion Zero Modes} on a soliton have played an important role 
in the study of quantum field theories and string/M theory.
To quantize them, we usually make pairs of them and define creation and annihilation operators.
The quantization of them gives the degeneracy to the soliton ground states and sometimes determines some of the quantum numbers of the soliton.
For example, the spin and the flavor charge of an 't Hooft-Polyakov monopole 
in an $SU(2)$ gauge theory are determined 
by quantizing the fermion zero modes around it; 
its analysis goes back to Jackiw and Rebbi \cite{Jackiw:1975fn}.

A more exotic situation arises when a soliton has \emph{an odd number of Majorana fermion zero modes}, where we cannot make pairs of them.
In $1+1$ and $2+1$ dimensions, having an odd number of Majorana fermion zero modes leads to non-abelian statistics of the states \cite{Kitaev:2001kla,Moore:1991ks}. Such objects are called anyons and are known to arise in topological superconductivity \cite{Read:1999fn}. The natural question is what happens when there are an odd number of Majorana fermion zero modes on solitons in higher dimensional theory.

\bigskip

\emph{In this thesis, we propose a new consistency condition of quantum field theories}: Having an odd number of Majorana fermion zero modes on a dynamical point-like soliton signifies an inconsistency in a theory with $3+1$ or higher dimensions. Given the example of the Witten anomaly, it is likely that the inconsistency is directly connected to the anomaly of discrete parts of the gauge symmetries.

\bigskip

The main application of our proposal is to check the \emph{consistency of string and M theory compactifications}. Anomalies of continuous gauge symmetries in string theory are proved to vanish \cite{Green:1984sg,Schellekens:1986xh,Lerche:1988np}. However, the complete understanding of the anomalies of discrete parts of gauge symmetries in string and M theory compactifications is still lacking. Our consistency condition gives one of the criteria for the problem. Solitons in string and M theories are realized by branes, and the fermion zero modes on them can be easily computed in the language of branes. We will discuss this issue in chapters \ref{chap:branes-wrap} and \ref{chap:branes-intersect}.

\bigskip

As a sub-issue, we discuss the \emph{realization of anyons by branes} in string and M theory compactifications in $2+1$ dimension. To the author's knowledge, the direct realization of anyons in string theory is not known yet. We will give a short comment on this issue in section \ref{sec:interpret-brane-intersect}.

\newpage

\section*{Outline}

The rest of this thesis is organized as follows.

Chapters \ref{chap:anom-QFT}, \ref{chap:anom-string} and \ref{chap:MFZ} are the review parts.
In chapter \ref{chap:anom-QFT}, we review aspects of anomalies in quantum field theory.
In chapter \ref{chap:anom-string}, we describe how the anomalies in string theories cancel.
In chapter \ref{chap:MFZ}, we introduce the solitons and the fermion zero modes in quantum field theory. In particular, we will see the relationship between the Witten anomaly and the fermion zero modes on an 't Hooft-Polyakov monopole.

We describe our main new results in chapters \ref{chap:consistency}, \ref{chap:branes-wrap}, and \ref{chap:branes-intersect}. 
In chapter \ref{chap:consistency}, we give the argument that an odd number of Majorana fermion zero modes on a point-like soliton signifies an inconsistency. 
In chapter \ref{chap:branes-wrap}, we discuss the fermion zero modes on a (coincident) M5-brane(s) wrapped around cycles.
In chapter \ref{chap:branes-intersect}, we discuss the fermion zero modes on intersecting D-branes.

In chapter \ref{chap:conclusion}, we conclude the thesis.

We describe the calculation methods appearing in this thesis in appendices \ref{app:open-spectrum}, \ref{app:fermions} and \ref{app:index-theorem}.
In appendix \ref{app:open-spectrum}, we describe the calculation of the spectrum on D-branes.
In appendix \ref{app:fermions}, we review the representation theory of spin groups.
In appendix \ref{app:index-theorem}, we summarize the index theorems used in this thesis.

\section*{Conventions}

We use the metric $\eta_{\mu\nu}=\text{diag}(-1,1,\cdots,1)$ for the Minkowski space $\bR^{1,d}$ and $\eta_{\mu\nu}=\text{diag}(1,\cdots,1)$ for the Euclidean space $\bR^d$. The descriptions of the dimensions of the spacetime are $d+1$ for Lorentzian and $d$ for Euclidean. One can see if the spacetime is Wick-rotated or not with these conventions. The Einstein summation should be taken if not specified. We use the metric $g_{\mu\nu}$, $g^{\mu\nu}:=(g^{-1})_{\mu\nu}$ of the spacetime to switch between the subscripts and the superscripts of spacetime tensors.

The gamma matrices $\Gamma^{\mu}$ of the spacetime satisfy $\{\Gamma^{\mu},\Gamma^{\nu}\}=2\eta^{\mu\nu}$. The chiral operator $\Gamma$ is defined to be
\begin{align}
	\Gamma:=c\prod_{\mu}\Gamma^{\mu}
\end{align}
where a constant $c$ is determined so that $(\Gamma)^2=1$. The Dirac conjugate of a fermion $\psi$ is $\overline{\psi}:=\psi^{\dagger}\Gamma^0$. See also appendix \ref{app:fermions} for fermions.

The representation matrices $T_i$ of a Lie algebra are normalized so that
\begin{align}
	\text{Tr}(T_iT_j)=\frac{1}{2}\delta_{ij}.
\end{align}
The characters $A$, $F$, and $R$ are usually used to describe the 1-form gauge field, the curvature 2-forms of the gauge group, and the spacetime curvature. The covariant derivative is $D:=D_{\mu}dx^{\mu}:=d+\rho(A)$ where $\rho$ is the representation of the gauge group and the Dirac operator is $\slashed{D}:=\Gamma^{\mu}D_{\mu}$. Sometimes we use subscripts to describe the rank of differential forms, such as $F_2$ for a 2-form $F$. The action of the group $G$ is described by the superscripts, such as $A^g:=g^{-1}Ag+g^{-1}dg$ for $g\in G$.

\subsection*{Special notes}

This thesis is based on the following papers

\begin{tabular}{ll}
	\cite{Sato:2022vii}
	& Y.\,Sato, Y.\,Tachikawa, T.\,Watari,``On odd number of fermion zero modes\\
	&on solitons in quantum field theory and string/M theory,''\\
	& \href{https://arxiv.org/abs/2205.13185}{
		\magenta{arXiv:2205.13185 [hep-th]}},
	JHEP\,\textbf{09} (2022) 043.\\
	\cite{ST:2024}
	& Y.\,Sato, Y.\,Tachikawa, ``On fermion zero modes at brane intersections,'' to appear.
\end{tabular}


\chapter{Anomalies in Quantum Field Theory}\label{chap:anom-QFT}


Let us begin with a review of anomalies in quantum field theory. We recommend \cite{Harvey:2005it} for a more detailed review. The textbook \cite{Nair:2005iw} might also be helpful. There is no complete textbook or review on modern perspectives of anomalies, but the papers \cite{Garcia-Etxebarria:2018ajm,Witten:2019bou} come close.

Suppose a \emph{classical} field theory with Lagrangian density $\cL(\phi)$ has a symmetry $G$ acting on the field $\phi$ as $G\ni g:\phi\rightarrow\phi^g$. The symmetry $G$ is called anomalous if it is violated in the \emph{quantum} field theory with the same Lagrangian density $\cL(\phi)$.

There are different kinds of symmetries in field theories: Global and gauge symmetries, continuous and discrete symmetries. 

The existence of the anomaly of a global symmetry indicates that the existence of non-trivial external gauge fields will break the symmetry in the quantum theory. Furthermore, it sometimes determines the forms of correlation functions.
There are applications of such anomalies to calculate the mass of the $\eta'$ meson for example. 
On the other hand, the existence of the anomaly of a gauge symmetry indicates the inconsistency of the quantum theory. 
We call it the gauge anomaly. 
Recall that in quantum gauge theory, a gauge symmetry is not a symmetry but a redundancy, and we need the gauge symmetry to fix the gauge and remove negative norm states from physical states. 
The lack of gauge symmetry leads to the lack of the unitarity of the theory. 
Therefore anomalies of gauge symmetries must vanish in the consistent quantum field theory. 
The cancellation of gauge anomalies is a non-trivial check of the consistency of the quantum field theory. 
For example, it is known that the gauge anomalies in the Standard Model vanish in a non-trivial way.

We call the anomalies of continuous symmetries perturbative anomalies because it suffices to analyze the infinitesimal transformations. The aspects of perturbative anomalies are reviewed in section \ref{sec:pert-QFT}. The anomaly of discrete symmetries\footnote{
	Here \emph{discrete symmetries} include the discrete parts of continuous symmetries. 
	An example is non-trivial component of $\pi_4(SU(2))=\bZ_2$, which we will later see in \ref{sec:non-pert-QFT}. 
	It is a part of continuous $SU(2)$ gauge symmetry but is not continuously connected to the identity. 
	The anomaly of such a component should be treated as a non-perturbative anomaly.
}, on the other hand, is harder to analyze in general. 
Such kind of anomaly is called non-perturbative anomaly in this thesis and is reviewed in section \ref{sec:non-pert-QFT}. 
In section \ref{sec:modern-QFT}, we give a brief review of the modern understanding of anomalies, which gives the mathematical classification of non-perturbative anomalies.

\section{Perturbative Anomalies}\label{sec:pert-QFT}
Let us see the perturbative anomaly of a continuous symmetry $G$. Classically, according to Noether's theorem, there is a conserved current $j^a_{\mu}$, where $a$ labels the generators of $G$, satisfying
\begin{align}
	\partial^{\mu}j^a_{\mu}=0.
\end{align}
The quantized theory does not have the same conserved current in general. The expectation value of $j^a_{\mu}$ could have non-trivial divergence
\begin{align}
	\partial^{\mu}\vev{j^a_{\mu}}=\cK^a.
\end{align}
The quantity $\cK^a$ is called the anomaly of this theory for the continuous symmetry $G$. The non-zero $\cK^a$ indicates that the symmetry $G$ in the classical theory is violated in the quantized theory.

\subsection*{Chiral Anomaly}

The simple and famous example is the chiral anomaly in $U(1)_V$ gauge theory in $1+1$-dimensional flat spacetime with the Lagrangian density
\begin{align}
	\cL=-\frac{1}{4e^2}F_{\mu\nu}F^{\mu\nu}+i\overline{\psi}\slashed{D}\psi,
\end{align}
where $F_{\mu\nu}$ is a field strength of the $U(1)_V$ gauge field and $\psi$ is a Dirac fermion with $U(1)_V$-charge 1.

Classically this theory is invariant under the global $U(1)_A$ chiral transformation $\psi\rightarrow e^{i\theta\Gamma}\psi$ where $\theta$ is a constant. The Noether current for this is
\begin{align}
	j_{\mu}=\overline{\psi}\Gamma_{\mu}\Gamma\psi.
\end{align}
The 1-loop calculation gives
\begin{align}\label{chiral-anom}
	\partial_{\mu}\vev{j^{\mu}}=\frac{1}{2\pi}\epsilon^{\mu\nu}F_{\mu\nu}.
\end{align}
This indicates that the $U(1)_A$ symmetry is violated for a non-zero $F_{\mu\nu}$.
Here, the renormalization respects the $U(1)_V$ symmetry. Note that the gauge field is fixed, or regarded as an external field so far. We need to path-integrate the gauge field later. 

This theory is consistent because the chiral $U(1)_A$ symmetry is global. Gauging the symmetry\footnote{
	To \emph{gauge} the symmetry, we usually add the gauge field coupled with charged fields and add the kinetic term of the gauge field to the Lagrangian.
}, however, leads to an inconsistency.

\subsection{Fujikawa method}

The Fujikawa method gives a simple interpretation of anomalies in quantum field theory. We use the path integral formula to compute the vacuum expectation value. The infinitesimal change of variables $\psi'=e^{i\alpha(x)\Gamma}\psi$ does not change the path integral, where $\alpha(x)$ is an arbitrary function. Therefore we get the relation
\begin{align}
	Z(A)&=\int D\psi D\overline{\psi}e^{i\int \cL(\psi,A)}=\int D\psi' D\overline{\psi'}e^{i\int \cL(\psi',A)}\\
	&=\int D\psi'D\overline{\psi'}e^{i\int\cL(\psi,A)}(1+\int dx^2\partial^{\mu}\alpha(x)j_{\mu}+\cO(\alpha^2)).
\end{align}
The third equation holds because $\psi'$ corresponds to the infinitesimal $U(1)_A$ transformation.

Suppose the Jacobian of this change of variables is trivial,
\begin{align}
	D\psi D\overline{\psi}=D\psi'D\overline{\psi'}.
\end{align}
Then, because $\alpha(x)$ is arbitrary, integrating by parts we get
\begin{align}
	\partial^{\mu}\vev{j_{\mu}}=0
\end{align}
which means that the anomaly vanishes. From this viewpoint, anomalies live in the measure of path integrals. Let us see that the careful calculation of the Jacobian reproduces the same anomaly as (\ref{chiral-anom}). The Jacobian for the transformation $\psi\rightarrow\psi'$ is
\begin{align}
	{\rm Det}(e^{-2i\alpha\Gamma})=1-2i{\rm Tr}(\alpha\Gamma)+\cO(\alpha^2)
\end{align}
where ${\rm Det}$ and ${\rm Tr}$ are in the functional space. The anomaly is
\begin{align}
	\int dx^2\alpha(x)\partial^{\mu}\vev{j_{\mu}}=-2i{\rm Tr}(\alpha\Gamma).
\end{align}

We need to regularize to compute the trace. We Wick-rotate the theory and work in the Euclidean signature. Respecting $U(1)_V$ symmetry, we choose $-\slashed{D}^2/M^2$ as the regulator:
\begin{align}\label{inf-sum}
	{\rm Tr}(\alpha\Gamma):={\rm Tr}(\alpha\Gamma e^{-\slashed{D}^2/M^2})={\rm Tr}(\alpha\Gamma e^{(-D^2+i\Gamma^{\mu\nu}F_{\mu\nu})/M^2})
\end{align}
where $\Gamma^{\mu\nu}:=[\Gamma^{\mu},\Gamma^{\nu}]/2$. Because this quantity does not depend on $M$\footnote{
	Note that states $\phi$ and $\slashed{D}\phi$ make a pair if $\slashed{D}\phi\neq0$. Because they have the same eigenvalues of $\slashed{D}^2$ and $\slashed{D}$ flips the eigenvalue of $\Gamma$, the contributions from them cancel. Therefore, only zero modes of $\slashed{D}$ contribute to the trace and the trace does not depend on $M$.
}, we take a limit $M\rightarrow \infty$. Then
\begin{align}\notag
	{\rm Tr}(\alpha\Gamma)&=\lim_{M\rightarrow\infty}\int dx^2\alpha(x){\rm tr}[\Gamma e^{-i\Gamma^{\mu\nu}F_{\mu\nu}/M^2}]\int \frac{dp^2}{(2\pi)^2}e^{-p^2/M^2}\\ \notag
	&=\frac{1}{4\pi}\int dx^2\alpha(x){\rm tr}[\Gamma(-i\Gamma^{\mu\nu}F_{\mu\nu})]\\ 
	&=\frac{i}{4\pi}\int dx^2\alpha(x)\epsilon^{\mu\nu}F_{\mu\nu}
\end{align}
which matches well with (\ref{chiral-anom}).

In general, the chiral $U(1)_A$ anomaly in a $(2n-1)+1$-dimensional spacetime $M$ is given by the index density
\begin{align}\label{chiral-anom-gen}
	\int_Mdx^{2n}\alpha(x)\partial^{\mu}\vev{j_{\mu}}=2\int_M\alpha(x)[\hat{A}(M)\wedge Ch(F)]_{2n}
\end{align}
for a Dirac fermion with charge 1 living in some representation of the symmetry group $G$. $F$ is the gauge bundle of $G$. $\hat{A}(M)=1+p_1(M)/24+\cdots$ is\footnote{
	$p_1(M)$ is called the first Pontrjagin class of $M$. As a differential form, it is
	\begin{align*}
		p_1(M)=-\frac{1}{8\pi^2}\text{tr}R^2.
	\end{align*}
} the $A$-roof genus of $M$ and $Ch(F)={\rm tr}e^{F/2\pi}$ is the Chern character of the gauge bundle. The bracket $[\ ]_n$ means that we take the $n$-form part of the differential form.
The origin of this formula comes from the fact that ${\rm Tr}(\Gamma)$ is equal to the difference between the number of the zero-eigenstates of $D$ and $D^{\dagger}$, which is equal to the integral of the index density. See appendix \ref{app:index-theorem} for the index theorem.

\subsection{Anomaly Polynomial}

We would like to generalize the above discussions to non-abelian symmetries. The anomalies in $(2n-1)+1$-dimensional theory is known to be controlled by $(2n+2)$-form $\cI_{2n+2}$ called the anomaly polynomial. Say we have a Weyl fermion $\psi$ in some representation of the symmetry group $G$ in the $(2n-1)+1$-dimensional spacetime $M$. The anomaly polynomial for this situation is given by
\begin{align}
	\cI_{2n+2}=[\hat{A}(M)\wedge Ch(F)]_{2n+2}.
\end{align}
Fermions with opposite chirality contribute by the minus of this.
The anomaly for the infinitesimal transformation $g=1+\alpha\in G$ is given by
\begin{align}
	\int_Mdx^{2n}\alpha\partial^{\mu}\vev{j_{\mu}}=2\pi \int_M\cA_n(\alpha,A)+\cO(\alpha^2)
\end{align}
where $\cA_n$ is defined so that
\begin{align}
	d\cA_n&=\delta_gCS_{2n+1},\\
	\cI_{2n+2}&=dCS_{2n+1}
\end{align}
for a $(2n+1)$-form $CS_{2n+1}$.

Let us calculate the anomaly using the anomaly polynomial for the example above, $U(1)_A$ chiral anomaly in $(2n-1)+1$-dimensional theory. We have a Weyl fermion with $U(1)_A$-charge $+1$, and an opposite Weyl fermion with $U(1)_A$-charge $-1$, both in some representation of the symmetry group $G$. The total symmetry group is $U(1)_A\times G$. We consider the infinitesimal transformation $g=1+\alpha\in U(1)_A$. The anomaly polynomial is
\begin{align}
	\cI_{2n+2}=2[\hat{A}(M)Ch(F_{U(1)_A\times G})]_{2n+2}=\frac{F_{U(1)_A}}{\pi}\wedge[\hat{A}(R)\wedge Ch(F_G)]_{2n}+\cdots.
\end{align}
The other terms are irrelevant. Then we get
\begin{align}
	CS_{2n+1}&=\frac{A}{\pi}[\hat{A}(M)\wedge Ch(F_G)]_{2n}+\cdots,\\
	\cA_n&=\frac{\alpha}{\pi}[\hat{A}(M)\wedge Ch(F_G)]_{2n}
\end{align}
which reproduces the formula (\ref{chiral-anom-gen}).

Perturbative anomalies in a theory vanish if and only if the relevant anomaly polynomials vanish. Therefore, the anomaly polynomials for gauge symmetries must vanish so that the theory is consistent. This is a strong tool to check the consistency of the theory. Furthermore, anomaly polynomials tell us if the theory has gravitational anomalies and mixed anomalies. Gravitational anomalies correspond to the terms including only the curvature $R$ of the spacetime $M$, and mixed anomalies correspond to the terms including both the curvature and the gauge field strength. The theory can consistently couple to gravity only if the gravitational and mixed anomalies vanish. In particular, in quantum gravity, all symmetries including gravity are believed to be gauged. Therefore the total anomaly polynomial must vanish in quantum gravity.

\subsection{Perturbative Anomalies in the Standard Model}

Since the Standard Model is a chiral gauge theory, the cancellation of perturbative gauge anomalies in the Standard Model is highly non-trivial. Let us compute the anomaly polynomial of the Standard Model and check the gauge anomaly cancellation.

The Standard Model has $U(1)\times SU(2)\times SU(3)$ gauge symmetry. Since the $SU(3)$ coupling is vectorial, we need to check the cancellation for the electroweak part. 
Weyl fermions in the Standard Model have 3 generations, and each of them comes with $(l_L,l_R,Q_L,U_R,D_R)$, where $R,L$ means right/left-handed. 
The charges of them are tabulated in Table \ref{table-standard-model}.

\begin{table}[h]
	\caption{Weyl fermions in the standard model}
	\label{table-standard-model}
    \centering
    \begin{tabular}{|c|c|c|c|}
        \hline
        &$U(1)$&$SU(2)$&$SU(3)$\\
		\hline
        $l_L$&$-1$&$\mathbf{2}$&$\mathbf{1}$\\
        $l_R$&$-2$&$\mathbf{1}$&$\mathbf{1}$\\
        $Q_L$&$1/3$&$\mathbf{2}$&$\mathbf{3}$\\
        $U_R$&$4/3$&$\mathbf{1}$&$\mathbf{3}$\\
		$D_R$&$-2/3$&$\mathbf{1}$&$\mathbf{3}$\\
        \hline
    \end{tabular}
\end{table}

The anomaly polynomial is
\begin{align}
	\cI_6=[\hat{A}(R)Ch(F)]_6=\frac{p_1(M)}{24}\frac{{\rm tr}(F)}{8\pi^2}+\frac{{\rm tr}(F^3)}{24\pi^3}.
\end{align}
First, let us check the cancellation of gauge anomalies ${\rm tr}F^3$. The pure $U(1)$ part contributes as
\begin{align}
	F_{U(1)}^3[2\cdot(-1)^3-(-2)^3+3\cdot2\cdot(1/3)^3-3\cdot(4/3)^3-3\cdot(-2/3)^3]=0.
\end{align}
Other terms that is not obviously traceless are ${\rm tr}F_{U(1)}{\rm tr}F_{SU(2)}^2$ and ${\rm tr}F_{U(1)}{\rm tr}F_{SU(3)}^2$. 
Note that other terms such as pure $SU(2)$ are traceless because of the tracelessness of the representation matrices. The contributions from the two terms are
\begin{align}
	F_{U(1)}F_{SU(2)}^2[(-1)+3\cdot(1/3)]&=0,\\
	F_{U(1)}F_{SU(3)}^2[2\cdot(1/3)-(4/3)-(-2/3)]&=0.
\end{align}
Therefore gauge anomalies are absent in the Standard Model.

Since the Standard Model is believed to be the low-energy effective theory of the quantum gravity in the real world, it is expected to consistently couple to gravity. The anomaly polynomial says the mixed anomaly is proportional to ${\rm tr}F$, which is further proportional to 
\begin{align}
	2\cdot(-1)-(-2)+3\cdot2\cdot1/3-3\cdot4/3-3\cdot(-2/3)=0.
\end{align} 
Therefore the mixed anomaly vanishes. There is no pure gravitational term. In the end, the Standard Model has miraculously no perturbative gauge, gravitational, and mixed anomalies.

\section{Non-Perturbative Anomalies}\label{sec:non-pert-QFT}

In this section we review the analysis of non-perturbative anomalies, which is the main interest in this thesis. Unfortunately, there is no unified way to compute non-perturbative anomalies. Later in section \ref{sec:modern-QFT} we will introduce the mathematical classification of non-perturbative anomalies, but its physical interpretation is another issue.

In this section, we just introduce one of the most famous and important examples, the $SU(2)$ Witten anomaly. One way to analyze it is to directly calculate the change of the partition function under the transformation. We will introduce another characteristic of non-perturbative anomalies in section \ref{sec:3+1-MFZ} and chapter \ref{chap:consistency}.

\subsection{Witten Anomaly}

The $SU(2)$ Witten anomaly is a non-perturbative anomaly in $3+1$-dimensional $SU(2)$ gauge theory with one doublet Weyl fermion $\psi$
\begin{align}
	\cL=\frac{1}{2g^2}{\rm tr}F_{\mu\nu}F^{\mu\nu}-i\overline{\psi}\slashed{D}\psi.
\end{align}
We have checked that the anomaly polynomial for a chiral $SU(2)$ doublet fermion vanishes in $3+1$ dimensions because of the tracelessness of the representation matrices. 
Therefore there is no perturbative gauge anomaly in this theory. 
However, the theory turns out to be anomalous under the gauge transformation that is not connected to the identity.

We work in the 4-dimensional Euclidean spacetime. We further compactify the spacetime to $S^4$, which is justified because the fields must vanish at infinity. The space of gauge transformations $\cG\ni g:S^4\rightarrow SU(2)$ of 4-dimensional $SU(2)$ gauge group has two components corresponding to the non-trivial fourth homotopy group of $SU(2)$: $\pi_4(SU(2))=\bZ_2$. We take $g_4\in\cG$ to be the representative of the non-trivial part of $\pi_4(SU(2))$. This is not connected to the identity transformation and cannot be captured by the analysis of the infinitesimal transformations. Therefore the cancellation of the anomaly under $g_4$, which is needed for the theory to be consistent, is not guaranteed only by the cancellation of the anomaly polynomial.

Let us calculate the change in the partition function before integrating out the gauge field
\begin{align}
	Z(A):=\int D\psi D\overline{\psi}e^{-\cL(\psi,A)}
\end{align}
following the original paper by Witten \cite{Witten:1982fp}. The relevant term is
\begin{align}
	\int D\psi D\overline{\psi}e^{i\overline{\psi}\slashed{D}\psi}=[{\rm Det}(i\slashed{D}(A))]^{1/2}.
\end{align}
This is because a doublet Weyl fermion is one-half of a doublet Dirac fermion. 

${\rm Det}(i\slashed{D})$ is formally defined by the infinite product of the eigenvalues of $i\slashed{D}$. If $\phi_{\lambda}$ is an eigenstate of $i\slashed{D}$ with the eigenvalue $\lambda$, $\Gamma \phi_{\lambda}$ is also an eigenstate with the eigenvalue $-\lambda$. Therefore the eigenvalues are paired up, and we get
\begin{align}
	{\rm Det}(i\slashed{D}(A))=\prod(\lambda(A))(-\lambda(A)).
\end{align}
Because $i\slashed{D}$ is Hermitean, $\lambda$'s are real. We choose so that all $\lambda$'s to be positive.

Taking the square root of this corresponds to choosing either $\lambda$ or $-\lambda$ for each pair,
\begin{align}
	[{\rm Det}(i\slashed{D}(A))]^{1/2}=\prod s_{\lambda}\lambda(A),\,\,\, s_{\lambda}=\pm1.
\end{align}
We can freely choose the signs, but the partition function must be smooth under the infinitesimal deformations once we fix the sign. 
So the definition of $[{\rm Det}(i\slashed{D})]^{1/2}$ has indeed invariant meaning under the gauge transformations connected to the identity because $\lambda(A)$'s are invariant. 
However, the invariance under $g_4$, which is not connected to the identity, is not guaranteed. We will show that $[{\rm Det}(i\slashed{D})]^{1/2}$ transforms as
\begin{align}\label{Witten-anom}
	[{\rm Det}(i\slashed{D}(A^{g_4}))]^{1/2}=-[{\rm Det}(i\slashed{D}(A))]^{1/2}.
\end{align}
This indicates that the partition function also transforms as $Z^{g_4}(A)=Z(A^{g_4})=-Z(A)$, and leads the inconsistency of the theory.

To show the equation (\ref{Witten-anom}), suppose we choose all $s_{\lambda}$'s to be $-1$.
Then consider the loop in the configuration space of the gauge field 
\begin{align}
	A(t)=(1-t)A+tA^{g_4}.
\end{align}
$[{\rm Det}(i\slashed{D})]^{1/2}$ is continuously deformed from $A$ along the loop coming back to $A^{g_4}$. Since $A$ and $A^{g_4}$ is gauge-equivalent, the sets of eigenvalues, $\{\lambda(A)\}$ and $\{\lambda(A^{g_4})\}$ are the same. However the sets of signs, $\{s_{\lambda(A)}\}$ and $\{s_{\lambda(A^{g_4})}\}$ could be changed. 
This is the origin of the sign in (\ref{Witten-anom}). 
If an odd number of $s_{\lambda}$ flips under the loop, the sign of the partition function flips and ends in the anomaly.

The number of flips is equal to the zero modes of 5-dimensional Dirac operator $D^5:=\Gamma d/dt+\slashed{D}(A_t)$ on $\bR_t\times S^4$, where $A_t$ is a 5-dimensional gauge field satisfying $A_t=A$ at $t\sim-\infty$ and $A_t=A^{g_4}$ at $t\sim\infty$. The reason is as follows. A zero mode $\psi$ of $D^5$ satisfies
\begin{align}
	d\psi/dt=-\Gamma\slashed{D}(A_t)\psi.
\end{align}
By changing the basis, $\Gamma\slashed{D}$ can be rewritten as $i\slashed{D}$.
Now expand $\psi=\sum\psi_a(t)\phi_a^t(x)$ by the eigenstates $\phi_a^t$ of $i\slashed{D}(A_t)$ with eigenvalue $\lambda_t$ depending on $t$ smoothly. Because $i\slashed{D}(A_t)$ is Hermitean, $\lambda_t$ are real. Each mode of $\psi$ satisfies
\begin{align}
	d\psi_a/dt=-\lambda_t\psi_a
\end{align}
which is solvable $\psi_a=e^{-\int dt\lambda_t}$. This is normalizable if and only if $\lambda_t<0$ at $t\sim-\infty$ and $\lambda_t>0$ at $t\sim\infty$. Therefore, a zero mode $\psi$ gives a smooth eigenstate of $i\slashed{D}(A_t)$ that changes the sign of the eigenvalue from negative to positive\footnote{
	Because the spectrum of $i\slashed{D}(A)$ and $i\slashed{D}(A^{g_4})$ are identical, it also gives a smooth eigenstates of $i\slashed{D}(A_t)$ that changes the sign from positive to negative. The choice of signs does not matter.
}, and vice versa.


The number of zero modes of $D^5$ can be calculated in principle by the mod 2 Atiyah-Singer index theorem \cite{Atiyah:1971rm}. See appendix \ref{sec:mod2-index} for the mod 2 index theorem. To the author's knowledge, no literature directly computes the index. Nevertheless, the number of zero modes is known to be 1 mod 2 from an abstract mathematical analysis. Therefore this theory is anomalous. In section \ref{sec:modern-QFT} one can find a justification that this theory indeed has a $\bZ_2$ anomaly. In section \ref{sec:3+1-MFZ} and chapter \ref{chap:consistency}, we will give another simple interpretation of this anomaly in terms of fermion zero modes on a monopole.

In general, the $Sp(N)$ gauge theory with an odd number of Weyl fermions in the representation $R$ with half-integral $C(R)$\footnote{
	$C(R)$ is the index of the representation $R$ and is defined by
	\begin{align*}
		{\rm tr}_R(T^aT^b)=C(R)\delta^{ab}
	\end{align*}
} is anomalous under the gauge transformation associated to $\pi_4(Sp(N))=\bZ_2$.

\subsection{Witten Anomaly in the Standard Model}

Since the Standard Model is a chiral $SU(2)$ gauge theory, it possibly has $SU(2)$ Witten anomaly. Chiral $SU(2)$ doublet fermions in the Standard Model are $Q_L$ and $l_L$ (See Table \ref{table-standard-model}). There are 4 Weyl doublet fermions in total in each generation, and the Witten anomaly cancels within each generation. Note that the Witten anomaly does not cancel within only leptons or only quarks. This is another miraculous anomaly cancellation in the Standard Model.

\subsection{Witten Anomaly in $4+1$ Dimensions}\label{subsec:4+1-non-pert-QFT}

Let us next consider an analogue in $4+1$ dimensions.
Namely, let us take an $SU(2)$ gauge theory
with $n$ fermions $\psi$ in the doublet with the symplectic-Majorana condition.
Here, the symplectic-Majorana condition on a fermion field $\psi_{a\alpha}$ in the doublet
is imposed by \begin{equation}
\psi_{a\alpha}= \epsilon_{ab}J_{\alpha\beta}(\psi^*)^{b\beta}
\end{equation}
where $a,b=1,2$ are the $SU(2)$ indices,
$\alpha,\beta=1,2,3,4$ are the spinor indices,
and $\epsilon_{ab}$ and $J_{\alpha\beta}$ are the antisymmetric invariant tensors
which exist because $\mathbf{2}$ of $SU(2)$ and $\mathbf{4}$ of $SO(4,1)$ are pseudo-real.

This theory is afflicted with a non-perturbative anomaly, 
this time associated with $\pi_5(SU(2))=\bZ_2$ instead of $\pi_4(SU(2))=\bZ_2$.

\section{Modern Perspective}\label{sec:modern-QFT}

Recently the theory of anomalies is understood in the language of the theory of bordisms. It is mathematically complicated and sometimes almost impossible to compute in practice. Also, it only gives the abstract classification of non-perturbative anomalies, and the physical interpretation is another issue to be addressed.

Nevertheless, it sheds light on the unified understanding of the non-perturbative anomalies in quantum field theories. 
Although the details of this perspective are not so related to our discussions later, we give some comments on bordisms and known results. The readers may skip this section and move to chapter \ref{chap:anom-string}.

\subsection{Eta Invariant and Bordisms}

We are interested in anomalies in a quantum field theory $\cQ(M)$ on an $n$-dimensional Euclidean compact spacetime $M$. Suppose there is a symmetry $G$ of $\cQ(M)$, possibly anomalous. $G$ can be any kind of symmetry: global or gauge, continuous or discrete. It is simply rephrased that $M$ is equipped with the $G$-bundle structure representing the gauge field.

We extend the theory $\cQ(M)$ to the theory on an $(n+1)$-dimensional manifold $N_0$ with the boundary $\partial N_0=M$. We glue two $N_0$'s by the transformation $g\in G$ and get the theory on the $(n+1)$-dimensional manifold $N$ without boundary. 
We will see below that the difference of the partition function $Z$ of $\cQ(M)$ can be written as
\begin{align}
	Z^g=e^{2\pi i\eta_N}Z.
\end{align}
$\eta_N$ is called the eta invariant of $N$, and formally defined to be
\begin{align}
	\eta_N=\frac{1}{2}\sum_{\lambda}s_{\lambda}
\end{align}
where $\lambda$'s are the eigenvalues of the $(n+1)$-dimensional massless twisted Dirac operator $\slashed{D}_N$ and $s_{\lambda}$ is the sign of $\lambda$\footnote{
	The sign of $\lambda=0$ is set to be $+1$. This formula is a generalization of the analysis of the Witten anomaly.
}. The anomaly of $\cQ(M)$ can be captured by the $(n+1)$-dimensional topological invariant $\eta_N$.

This formula is justified as follows. We have seen that chiral fermions on $M$ are the sources of anomalies\footnote{
	When $\cQ(M)$ has Majorana-Weyl fermions, the formula will have an additional $1/2$.
}. A chiral massless fermion on $M$ is realized as the boundary state of a massive Dirac fermion on the bulk $N_0$. The regularized partition function of the massive fermion is 
\begin{align}
	Z_{N_0}=\prod_{\lambda}\frac{\lambda+im}{\lambda-iM}.
\end{align}
The bare mass and the regularization mass do not affect the conclusion, so we take $m,M\rightarrow\infty$. With this limit the phase part of $Z_{N_0}$ is equal to $2\pi i\eta_{N_0}$. The anomaly of $\cQ(M)$ is understood as the anomaly inflow from the bulk. Therefore $\cQ(M)$ should have the same phase as $Z_{N_0}$. Now the phase of $Z^g/Z$ is equal to the difference $2\pi i (\eta_{-N_0^g}-\eta_{N_0})$, which is equal to $2\pi i\eta_N$ by the gluing theorem of $\eta$. See \cite{Witten:2019bou} for more rigorous and detailed discussions.

Our goal is to classify the possible $\eta_N$ of the theory $\cQ(M)$. Suppose that $N$ is also a boundary of an $(n+2)$-dimensional manifold $W$. Then $\eta_N$ can be calculated by the Atiyah-Patodi-Singer index theorem \cite{Atiyah:1975jf}
\begin{align}
	{\rm ind}D_W=\eta_N+\int_W\hat{A}(R)Ch(F)
\end{align}
where $D_W$ is the Dirac operator on $W$ and ${\rm ind}D_W$ is the difference of the number of zero modes of $D_W$. See also appendix \ref{sec:APS-index} for the Atiyah-Patodi-Singer index theorem. ${\rm ind}D_W$ does not contribute to the anomaly because it is an integer. The integral part has exactly the form of the anomaly polynomial. This is the origin of the anomaly polynomial of perturbative anomalies from the modern point of view.

Now suppose the anomaly polynomial of the theory $\cQ(M)$ vanishes. One might think $\eta$ does not contribute to the phase of the partition function anymore. However, it may happen that there is no $(n+2)$-dimensional manifold $W$ satisfying $\partial W=N$ if $g$ is not connected to the identity. In such case we cannot use the Atiyah-Patodi-Singer index theorem and $\eta$ may produce a non-trivial phase. This is the origin of non-perturbative anomalies.

To classify the possible $\eta$ assuming the vanishing of the anomaly polynomials, note that two $(n+1)$-dimensional manifolds $N$, $N'$ have the same $\eta$ mod $\bZ$ if they are bounded by an $(n+2$)-dimensional manifold $W'$ satisfying $\partial W'=N\sqcup N'$. So we identify $N$ and $N'$. The set of $(n+1)$-dimensional spin manifold with $G$-bundle up to such identification is denoted by
\begin{align}
	\Omega_{n+1}^{spin}(BG)
\end{align}
and is called the bordism group. It is a group under the disjoint union $\sqcup$. 
Therefore, classification of possible $\eta$ is equal to computing $\Omega_{n+1}^{spin}(BG)$. In other words, non-perturbative anomalies of $\cQ(M)$ under $G$ are classified by the bordism group, and $\eta$ gives a map from the bordism group to the change of the phase $U(1)$.

\subsection{Example}

Some bordism groups were computed by mathematicians. See e.g. \cite{Wan:2018bns}. For example, the bordism groups for the 4-dimensional $SU(N)$ gauge theories with fermions are
\begin{align}
	\Omega_5^{spin}(BSU(2))&=\bZ_2,\\
	\Omega_5^{spin}(BSU(N))&=0,\,\,\,\text{for }N\geq3.
\end{align}
This indicates that $SU(2)$ theory has $\bZ_2$-valued non-perturbative anomaly, whereas $SU(N)$, $N\geq3$ do not have non-perturbative anomaly.
The non-trivial element of $\Omega_5^{spin}(BSU(2))$ corresponds to the Witten anomaly discussed above. It is indeed $\bZ_2$-valued. We also see that there is no analog of the $SU(2)$ Witten anomaly for $SU(N)$, $N\geq3$.


\chapter{Anomalies in String Theory}\label{chap:anom-string}


Since the string theories, or the low-energy effective super gravitational theories, are quantum gravitational theories, the gauge, gravitational, and mixed anomalies must vanish. In this chapter, we will review how the anomalies in string theories are canceled. The readers can skip this chapter and immediately move to chapter \ref{chap:MFZ}, keeping in mind that non-perturbative anomalies in string theories are not fully understood yet, whereas perturbative anomalies are well understood.

In section \ref{sec:sugra-string} we discuss the cancellation of perturbative anomalies in the smooth compactifications of low-energy effective supergravity. In section \ref{sec:pert-string}, we discuss the cancellation of perturbative anomalies in string theories in terms of the internal conformal field theory (CFT), and the effects of D-branes.
We also comment on the non-perturbative anomalies in string theories in section \ref{sec:non-pert-string}. There are only item-by-item discussions for non-perturbative anomalies in string theories, and we do not have the whole picture of them.

\section{Perturbative Anomalies in Supergravity}\label{sec:sugra-string}

First, we check the cancellation of the perturbative gauge, gravitational, and mixed anomalies in the smooth compactifications of low-energy effective supergravity theory. Since anomalies are the infrared effects, the cancellation of anomalies in supergravity is equal to the cancellation of anomalies in smooth compactifications of string theories.
In the previous chapter we have seen that chiral fields are the origin of the perturbative anomalies. Immediately we see that type IIA supergravity has no perturbative anomalies, because it is a non-chiral theory. On the other hand, type IIB and type I supergravities are possibly anomalous because they are chiral theories. See e.g. \cite{Polchinski:1998rr,Green:1984sg} for the perturbative anomalies in supergravity.

The chiral fields arising in supergravities are the Majorana-Weyl (MW) fermions, the MW gravitini, and the self-dual tensor fields in $9+1$ dimensions. The anomaly polynomial for a MW fermion in the representation $E$ was
\begin{align}\notag
	\cI_{12}^{MW}=&\frac{1}{2}[\hat{A}(M)Ch(F)]_{12}\\ \notag
	=&-\frac{{\rm tr}F^6}{1440}+\frac{{\rm tr}F^4{\rm tr}R^2}{2304}-\frac{{\rm tr}F^2{\rm tr}R^4}{23040}-\frac{{\rm tr}F^2({\rm tr}R^2)^2}{18432}\\ \label{anom-poly-MW10}
	&+{\rm dim}E\frac{{\rm tr}R^6}{725760}+{\rm dim}E\frac{{\rm tr}R^4{\rm tr}R^2}{552960}+{\rm dim}E\frac{({\rm tr}R^2)^3}{1327104}.
\end{align}
The coefficient $1/2$ in the first equality comes from the Majorana condition. We have redefined the field strength so that $F=F_{old}/2\pi$ for simplicity. We assumed ${\rm tr}F=0$.

The anomaly polynomials of a gravitino and a self-dual tensor can be deduced from (\ref{anom-poly-MW10}). A gravitino field is made from a left-moving vector and a right-moving MW fermion in the internal CFT. The product representation of them makes a MW gravitino $\mathbf{56}$ and a MW fermion $\mathbf{8}'$. Therefore, the anomaly polynomial for a MW gravitino is equal to
\begin{align}\notag
	\cI_{12}^{MWg}&=\cI^{MW}_{12}|_{F\rightarrow R, {\rm dim}E\rightarrow 8}+\frac{1}{2}[\hat{A}(M)]_{12}\\
	&=-495\frac{{\rm tr}R^6}{725760}+225\frac{{\rm tr}R^4{\rm tr}R^2}{552960}-63\frac{({\rm tr}R^2)^3}{1327104}.
\end{align}
A self-dual tensor is made from MW fermions $\mathbf{8}$ from each mover. Therefore the anomaly polynomial is
\begin{align}\notag
	\cI_{12}^{SD}&=-2\cI^{MW}_{12}|_{F\rightarrow R,{\rm dim}E\rightarrow8}\\
	&=992\frac{{\rm tr}R^6}{725760}-448\frac{{\rm tr}R^4{\rm tr}R^2}{552960}+128\frac{({\rm tr}R^2)^3}{1327104}
\end{align}
where $MW(M)$ is the bundle of MW fermion over $M$. The coefficient $-2$ in the first equality is needed because the field is a boson\footnote{
	Bosonic fields have $-2$ contributions compared to fermions. This is clear because the kinetic term of a boson contributes by ${\rm Det}(\slashed{D})^{-2}$ to the partition function, whereas a fermion contributes by ${\rm Det}\slashed{D}$.
}.

\subsection*{typeIIB}

Let us calculate the anomaly polynomial of the type IIB supergravity. Type IIB supergravity has two MW fermions $\mathbf{8}'$, two MW gravitini $\mathbf{56}$, and a self-dual tensor $[4]_+$. Therefore the total anomaly polynomial is
\begin{align}
	\cI_{12}^{\rm IIB}=-2\cI_{12}^{MW}|_{F\rightarrow0,{\rm dim}E\rightarrow1}+2\cI_{12}^{MWg}+\cI_{12}^{SD}=0.
\end{align}
In the end, we have checked that type IIB supergravity has no perturbative gravitational anomaly.

\subsection*{type I}

Type I and heterotic strings have the same effective low-energy type I supergravity. It has a MW fermion $\mathbf{8}'$, a gravitino $\mathbf{56}$ and a MW fermion $\mathbf{8}$ in the adjoint representation of the gauge group $G$. The total anomaly polynomial is
\begin{align}\notag
	\cI_{12}^G=&\frac{1}{1440}\left[-{\rm tr}F^6+\frac{{\rm tr}F^2{\rm tr}F^4}{48}-\frac{({\rm tr}F^2)^3}{14400}\right]\\
	&+({\rm dim}E-496)\left[\frac{{\rm tr}R^6}{725760}+\frac{{\rm tr}R^4{\rm tr R^2}}{552960}+\frac{({\rm tr}R^2)^3}{1327104}\right]+Y_4X_8
\end{align}
where
\begin{align}
	Y_4={\rm tr}R^2-\frac{1}{30}{\rm tr}F^2
\end{align}
and $X_8(R,F)$ is an 8-form. The first 2 terms in the anomaly polynomial vanishes for $G=SO(32),E_8\times E_8$. The last term $Y_4X_8$ is cancelled by the tree-level contribution from the higher derivative counter term
\begin{align}\label{GS-action}
	S_{GS}=\int B_2\wedge X_8(R,F).
\end{align}
Recall that in supergravity the field strength $H_3$ of the $B_2$-field has additional Chern-Simons terms
\begin{align}
	H_3=dB_2-c_A\omega^A_3-c_R\omega^R_3
\end{align}
where $c_A,c_R$ are constants and $\omega^A_3,\omega^R_3$ are the Chern-Simons 3-forms satisfying
\begin{align}
	d\omega^A_3={\rm tr}F^2,\,\,\, d\omega^R_3={\rm tr}R^2.
\end{align}
$B_2$ must be transformed under the gauge and gravitational transformations so that $H_3$ is invariant. Then the contribution of the term $S_{GS}$ to the anomaly polynomial is
\begin{align}
	(c_A{\rm tr}F^2+c_R{\rm tr}R^2)\wedge X_8.
\end{align}
The constants $c_A,c_R$ are determined by the string amplitudes, and they are exactly such that the anomaly polynomial cancels.
This is called the Green-Schwarz mechanism. In the end, type I supergravity has no perturbative gauge, gravitational, and mixed anomalies.

\bigskip

We have checked that the smooth compactifications of the low-energy effective supergravities have no perturbative anomalies. It means that the smooth compactifications of string theories also have no perturbative anomalies. However, the low-energy effective theories of the vacua of string theories are not necessarily of the form of smooth supergravities. We will discuss a more general class of string vacua in the next section.

\newpage

\section{Perturbative Anomalies in String Theory}\label{sec:pert-string}

We have confirmed that the smooth compactifications of string theories do not have perturbative gauge, gravitational, and mixed anomalies. In this section, we discuss more general string vacua in terms of internal worldsheet conformal field theories (CFTs). We also discuss the anomaly cancellation of string vacua with branes.


\subsection{In terms of CFT}

The perturbative anomalies in vacua of Heterotic and type II string theories without branes are known to cancel because of the modularity of internal conformal field theories \cite{Schellekens:1986xh,Lerche:1988np}. 
Following the original papers, we will show that an internal heterotic conformal field theory gives a spacetime theory with no perturbative anomalies.
The analysis of type II theories is essentially the same.

The internal heterotic CFT is made of right-moving supersymmetric CFT and left-moving bosonic current algebra CFT with central charges $(c,\tilde{c})=(15,26)$. The $(2n+2)$ non-compact directions of the CFT correspond to the spacetime in the string theory.

Let us define a character-valued partition function of the internal CFT
\begin{align}
	Z(\lambda,\tau,\overline{\tau}):={\rm Tr}(e^{2\pi i\lambda\cdot J}q^{H_L}\overline{q}^{H_R})
\end{align}
where $q=e^{2\pi i\tau}$, $H_{L/R}$ are the Hamiltonians for the left/right-mover, and $J$ is the set of generators of the symmetry of the theory including the left-moving gauge charges and the spacetime Lorentz charges of the states. 
The trace is given by the torus partition functions, and the trace is decomposed into the sum of four terms corresponding to the spin structure of the torus $(\pm,\pm)$. 
We focus on the $(+,+)$ term $Z_{++}$. In this term the right-moving ghost contributions cancel. We set the parameters $\lambda$ to be $x_{\alpha}$ for the Lorentz parts, and $\nu_{\alpha}$ for the gauge part, where $x_{\alpha}$, $\alpha=1,\cdots,n+1$, are the eigenvalues of the Riemannian tensor and $\nu_{\alpha}$ are set to be the eigenvalues of the gauge field strength. As described in detail in \cite{Schellekens:1986xh,Lerche:1988np}, we get
\begin{align}
	Z_{++}(\lambda,\tau,\overline{\tau})=\left(\prod_{\alpha}\frac{x_{\alpha}}{2\pi}\right)\cA(q,F,R).
\end{align}
$\cA(q,F,R)$ is defined as
\begin{align}
	\cA(q,F,R)=\hat{A}(R)\sum_{(g,r)}\epsilon(g,r)q^{m_L}Ch(F_g)Ch(R_r)
\end{align}
where the sum is over the left-moving states with gauge and Lorentz charges $(g,r)$ that is tensored by the right-moving ground states with chirality $\epsilon(g,r)$. $m_L$ is the left-moving mass term. Clearly, the $(2n+4)$-form part of the coefficient of $q^0$ in the $\cA(q,F,R)$ is equal to the anomaly polynomial of the spacetime theory.

We can determine the modular transformation property of $\cA(q,F,R)$ by the modular transformation property of the partition function $Z(\lambda,\tau,\overline{\tau})$. We get
\begin{align}
	\cA(e^{2\pi i\frac{a\tau+b}{c\tau+d}},\frac{F}{c\tau+d},\frac{R}{c\tau+d})=(c\tau+d)^{-n}{\rm exp}\left[\frac{ic}{32\pi^3(c\tau+d)}({\rm tr}F^2-{\rm tr}R^2)\right]\cA(q,F,R).
\end{align}
It means that
\begin{align}
	\cB(q,F,R):={\rm exp}\left[\frac{E_2(\tau)}{64\pi^2}({\rm tr}F^2-{\rm tr}R^2)\right]\cA(q,F,R)
\end{align}
transforms with modular weight $-n$, where $E_2(\tau)=1/3+\cO(q)$ is the Eisenstein series of weight 2, because $E_2(\tau)$ has the modular behavior of the form
\begin{align}
	E_2(\frac{a\tau+b}{c\tau+d})=(c\tau+d)^2E_2-\frac{2ic}{\pi}(c\tau+d).
\end{align}
Now let $\cB_{2n+4}(q)$ be the $(2n+4)$-form part of $\cB(q,F,R)$. As a function of $\tau$, $\cB_{2n+4}(q)$ transforms with modular weight 2, and is holomorphic except at infinity. 
From the general theory of modular forms, the coefficient of $q^0$ in the Fourier expansion of such a function is known to be zero. Consequently, the $(2n+4)$-form part of the coefficient of $q^0$ in $\cB(q,F,R)$ is zero. Immediately it means the $(2n+4)$-form part of the coefficient of $q^0$ in $\cA(q,F,R)$ must be of the form
\begin{align}
	[\cA(q,F,R)|_{q_0}]_{2n+4}=({\rm tr}F^2-{\rm tr}R^2)\wedge X_{2n}.
\end{align}
Recall that this is equal to the anomaly polynomial of the theory. This has a form that can be canceled by the Green-Schwarz mechanism. In the end, the perturbative anomalies are absent in any vacua in the heterotic string without branes.

\subsection{Anomaly Inflow of D-branes}

Consider a vacuum of type IIB string with a D$p$-brane, $p$ odd.
A worldvolume theory of the D$p$-brane possibly has Weyl fermions charged under the gauge and gravitational transformations. Such fermions contribute to the perturbative anomaly by $[\hat{A}(R)Ch(F)]_{p+3}$. Such anomaly is canceled by the anomaly from the bulk theory \cite{Green:1996dd}. It is called the anomaly inflow mechanism of branes. 

Let $M$ be the $d+1$-dimensional spacetime of the theory, and let there be a D$p$-brane in the spacetime. The bulk type IIB theory has a ($p+1$)-form $C_{p+1}$ coupling to the D$p$-brane. The D$p$-brane is a magnetic source of $C_{p+1}$:
\begin{align}
	d*H_{p+2}=:d\tilde{H}_{d-p-1}=\delta(W)
\end{align}
where $H_{p+2}$ is the field strength of $C_{p+1}$ and $\delta(W)$ is the delta function supported by the worldvolume of the D$p$-brane. The action of the bulk theory has the corresponding Chern-Simons term
\begin{align}
	S_{CS}=\int_M\tilde{H}\wedge\omega(R,F)_{p+2}
\end{align}
where $\omega(R,F)_{p+2}$ is the Chern-Simons $(p+2)$-form satisfying
\begin{align}
	d\omega(R,F)_{p+2}=[\hat{A}(R)Ch(F)]_{p+3}.
\end{align}
The term $S_{CS}$ contributes to the anomaly polynomial by $-[\hat{A}(R)Ch(F)]_{p+3}$ and cancels the anomaly from D$p$-brane. The theory is consistent in total.

The anomaly polynomial for the theory on M5-branes in M-theory compactifications can also be computed from the inflow mechanism \cite{Witten:1996hc}. We will use the result to compute the number of fermion zero modes on M5-branes in section \ref{sec:geom-branes-wrap}.

\section{Non-Perturbative Anomalies in String Theory}\label{sec:non-pert-string}

We confirmed that string theories do not have perturbative anomalies. On the other hand, we have little knowledge about the non-perturbative anomalies in string theories. The understanding of non-perturbative anomalies is crucial to check the consistency of string theory because all symmetries, including discrete symmetries, are believed to be gauged in quantum gravity and the anomalies of them must vanish.

There are only item-by-item discussions in the literature. Some of them assume geometric phases, others are applied to limited situations, or they are only conjectural. Let us review some of them.

\subsection*{Heterotic String}

Witten and Stong \cite{Witten:1985bt} pointed out that the non-perturbative anomalies in $E_8\times E_8$ heterotic string theory are classified by the bordism group
\begin{align}
	\Omega_{11}^{spin}(BE_8),
\end{align}
and it is proved to be zero. It indicates that $9+1$-dimensional $E_8\times E_8$ heterotic string theory has no non-perturbative anomaly. Further, the smooth geometric compactifications of it also have no non-perturbative anomaly.

From the standpoint that string vacua are given by the internal CFT possibly with branes, however, it should not be the whole story. There are other works \cite{Tachikawa:2021mby,Yonekura:2022reu} showing the non-perturbative anomaly cancellation of heterotic strings by using the topological modular forms of the internal CFT. Such approaches fully use the equivalence of the space of SCFT and the space of topological modular forms, which is conjectural so far \cite{Stolz:2011zj}.

There is also an approach to investigate the non-perturbative anomaly by using the classification of $3+1$-dimensional $\cN=2$ supersymmetric heterotic vacua by vector-valued modular forms \cite{Enoki:2020wel} written by the author. They discuss the Witten anomaly of the theory when $SU(2)$ gauge symmetry is enhanced. 
However, the cancellation of the Witten anomaly is not visible only from the known consistency conditions of string theories that appeared in the paper.

\subsection*{With Branes}

An approach to treat the D-branes is to classify them by geometric KO-theory. Freed \cite{Freed:2000ta} showed that type I theory with or without D-branes has no non-perturbative anomaly by using the geometric KO-theory. Again this discussion only holds for the geometric phase, not the general CFT. There is also a remaining question of the anomalies from intersecting D-branes.


\chapter{Solitons and Majorana Fermion Zero Modes}\label{chap:MFZ}


Solitons play important roles in understanding the non-perturbative structure of quantum field theories. In this chapter, we will review solitons in quantum field theory and fermion zero modes on them.

The definition of solitons depends on the literature. Here we simply define that solitons are the finite-energy nontrivial stable solutions of classical equations of motion. Typically the stability of the configuration is protected by topology; \emph{A soliton has a topological feature.} Because of the finiteness of the energy, soliton configurations should be localized at a region; \emph{A soliton is point-like}\footnote{
	Sometimes we consider objects with spatial dimensions greater than 0 such as branes. We do not call them solitons because they have infinite mass, but compactifications will make them finite energy. We consider such situations in chapters \ref{chap:branes-wrap} and \ref{chap:branes-intersect}.
}.

The statistics of solitons are sometimes determined by the quantization of fermion zero modes localized on the soliton. 
In this chapter, we introduce some kinds of solitons and discuss the quantization of fermion zero modes around them. 
In section \ref{sec:low-MFZ}, we analyze $1+1$ and $2+1$-dimensional models. 
We see that fermion zero modes on a soliton determine some quantum number of the soliton. Further, we see that an odd number of Majorana fermion zero modes leads to the non-abelian statistics of the states. We also comment on the relation between the fermion zero modes and superconductors. 
In section \ref{sec:3+1-MFZ}, we discuss the higher dimensional cases. It turns out that an odd number of fermion zero modes signifies an inconsistency.


\section{Low Dimensional Cases}\label{sec:low-MFZ}

\subsubsection*{Kink in $1+1$ Dimensions}

Our first example is a kink in $1+1$-dimensional theory with the Lagrangian density
\begin{align}
	\cL=\partial_{\mu}\phi\partial^{\mu}\phi-V(\phi)
\end{align}
where $\phi$ is a real scalar and $V(\phi)$ is a double-well potential whose minima are at $\phi=\pm \phi_0$. 

In this model, there is a dynamical domain wall (kink) connecting two vacua. This model is solvable if we set $V=(\phi^2-\phi_0^2)^2/2$ for example. The solution is $\phi=\pm\phi_0$tanh$(\phi_0x)$. The energy density of this configuration is $\cE=(\partial_x\phi)^2+V(\phi)$. This vanishes at infinity and the total energy is finite indeed. This soliton is associated to the topological invariant $\pi_{d-1}(C)=\pi_0(\bZ_2)=\bZ_2$ where $C$ is the configuration space of vacua.

\subsubsection*{Vortex in $2+1$ Dimensions}

Next, let us consider the $2+1$-dimensional theory with a complex scalar $\phi$ and a potential $V(\phi)$ whose minima are at $|\phi|=\phi_0\neq0$. The vortex solutions are associated with $\pi_1(C)=\bZ$, and classified by winding numbers. The typical form of the vortex solution with winding number $n$ is
\begin{align}
	\phi(r,\theta)\sim \phi_0e^{in\theta}
\end{align}
around $r\sim\infty$.

\subsection{Fermion Zero Modes}

When a theory has fermions coupled with gauge fields and scalars, there will be massless fermions in a soliton background localized around the soliton. We will call them fermion zero modes on the soliton. Quantizing the fermion zero modes, the creation operators act on the soliton ground states and form a local Hilbert space. 
Since the fermions are massless, they do not change the energy of the states. 
As a result, the ground states become degenerate. Such creation operators assign some quantum number to the soliton. The analysis of fermion zero modes goes back to \cite{Jackiw:1975fn}. One can also find a nice review in section 4 of \cite{Harvey:1996ur}.

Following the analysis in \cite{Jackiw:1975fn}, we add a Dirac fermion to the previous $1+1$-dimensional model with kinks. The Lagrangian density is
\begin{align}
	\cL=\partial_{\mu}\phi\partial^{\mu}\phi+\overline{\psi}\slashed{\partial}\psi+\phi\overline{\psi}\psi-V(\phi)
\end{align}
where $\psi$ is a Dirac fermion. It is easy to see that there is a single complex fermion zero mode in a kink background\footnote{
	Decomposing the fermion as $\psi=(\psi_+,\psi_-)^T$, the fermion zero modes are the solutions of the static Dirac equation
	\begin{align*}
		\partial\psi_{\pm}=\pm\phi\psi_{\pm}.
	\end{align*}
	We solve these equations for $\phi=$tanh$(x)$. The solution is
	\begin{align*}
		\psi_{\pm}=C(e^x+e^{-x})^{\pm 1}
	\end{align*}
	for a complex constant $C$. The fermion zero modes should be normalizable, so only $\psi_-$ is physical.
}. 

Therefore we have creation/annihilation operators $a^{\dagger},a$ satisfying
\begin{align}
	\{a^{\dagger},a\}=1,\{a,a\}=\{a^{\dagger},a^{\dagger}\}=0
\end{align}
acting on the kink ground state. 
As a result, the kink ground states are degenerate;
\begin{align}
	\ket{\Omega},a^{\dagger}\ket{\Omega}
\end{align}
where $\ket{\Omega}$ is a kink ground state satisfying $a\ket{\Omega}=0$. $a^{\dagger}$ increases the fermion number by 1. The two states are interpreted to be the kink and the anti-kink states, with fermion numbers $\pm1/2$. The assignment of degenerate local Hilbert space and quantum numbers are common phenomena of quantizing fermion zero modes on a soliton.

\subsection{Odd Number of Majorana Fermion Zero Modes}\label{subsec:odd-MFZ}

We have studied that fermion zero modes assign a local Hilbert space on the monopole ground states and determine some quantum number of them. 
However, the argument is applicable only if we have an even number of Majorana (real) fermion zero modes because the construction of annihilation/creation operators requires an even number of them. Models with an odd number of Majorana fermion zero modes on a soliton are known to exist, and their importance has been recognized recently.

\subsubsection*{Kink in $1+1$ Dimensions}

Instead of adding a Dirac fermion to the $1+1$-dimensional model with kinks, adding a Majorana fermion leads to a single Majorana fermion zero mode in a kink background.
When the scalar $\phi$ is made non-dynamical and the space is made discrete, the model reduces to Kitaev's quantum wire \cite{Kitaev:2001kla}.

This makes it impossible to assign a local Hilbert space associated with a single kink.
This is due to the following:
A well-separated pair of a kink and an anti-kink has two nearly-degenerate ground states,
coming from the quantization of two Majorana fermion zero modes.
If a local Hilbert space $\mathcal{H}$ can be assigned to each of the solitons, 
its dimension has to satisfy $(\dim\mathcal{H})^2=2$, making $\dim\mathcal{H}=\sqrt2$, 
which is impossible.
This is a symptom of non-abelian statistics of the state, on which we will comment later.

\subsubsection*{Vortex in $2+1$ Dimensions}

Recall that vortices are the solitons in $2+1$-dimensional theory with a complex scalar $\phi$ associated with $\pi_1(C)=\bZ$.
We add a Dirac fermion $\psi$ which couples to the scalar by the Yukawa coupling $\phi\psi\psi+cc.$ It is known that there is a single Majorana fermion zero mode on the vortex with a minimal winding number. 

This is a classic result going back to \cite{Jackiw:1981ee,Weinberg:1981eu}, whose importance in the physics of topological superconductors was recognized more recently. As in the case of the kink in $1+1$ dimensions, a single Majorana fermion zero mode on a vortex would assign a local Hilbert space of the dimension $\sqrt{2}$.

\subsection{Non-abelian Statistics and Superconductors}\label{subsec:superconductor-soliton}

We take the vortex in $2+1$ dimension as an example and see what happens with an odd number of Majorana fermion zero modes.
A nice summary can be found in section 3 and in Appendix B of \cite{Seiberg:2016rsg}.
As discussed there, the presence of an odd number of Majorana fermion zero modes leads to the non-abelian statistics possessed by the vortices\footnote{%
This observation goes back to \cite{Moore:1991ks}.
}.

Consider a configuration with two isolated vortices with minimal winding numbers in the 2-dimensional space. 
There is a single Majorana fermion zero mode on each of them. The action $\psi_i$, $i=1,2$ of each Majorana fermion zero mode on the ground states obeys the Clifford algebra
\begin{align}
	\{\psi_i,\psi_j\}=2\delta_{ij}.
\end{align}

Then let us define an operator $P$ that exchanges the two vortices. 
It must exchange also the fermion zero modes, and commute with the fermion parity $(-1)^F=i\psi_1\psi_2$. The operation
\begin{align}
	P:(\psi_1,\psi_2)\rightarrow(\psi_2,-\psi_1)
\end{align}
satisfies the conditions. 
Note that the change of the sign is needed to commute with $(-1)^F$. Due to the change of the sign, $P$ is no longer of order 2 but order 4: $P^2=-1$, $P^4=1$. 

In general, when there are $2N$ vortices, the ground states form a $2^N$-dimensional spinor representation of the fermion zero modes $\psi_i$, $i=1,\cdots,2N$. The actual form of the operation $P_{ij}$ that exchanges $i$-th and $j$-th vortices can be expressed by the matrix
\begin{align}
	P_{ij}=\pm\frac{1+\psi_i\psi_j}{\sqrt{2}}.
\end{align}
It indeed satisfies $P_{ij}^{-1}(\cdots,\psi_i,\psi_j,\cdots)P_{ij}=(\cdots,\psi_j,-\psi_i,\cdots)$. Note that they do not commute in general: The two operators $P_{ij}$ and $P_{jk}$ satisfy
\begin{align}
	[P_{ij},P_{jk}]=[\psi_i,\psi_k].
\end{align}
Having such statistics under the exchange indicates that the vortex has non-abelian statistics, neither bosonic nor fermionic. Such a particle-like state is called a non-abelian anyon.

Anyons play a central role in the theory of topological superconductivity.
The existence of anyons in topological superconductors was pointed out decades ago \cite{Read:1999fn}. 
Furthermore, since the anyonic states are robust against noises, they are possibly applied to topological quantum computations. 
One can find nice reviews in \cite{Sarma:2015pga,Masaki:2023rtn}. 
These applications to condensed matter physics motivate us to investigate the possibility of anyons in string theories. We will give comments on the problem in chapter \ref{chap:branes-intersect}.

\newpage

\section{$3+1$ or Higher Dimensions}\label{sec:3+1-MFZ}

Next, we will see the fermion zero modes on a soliton in a theory with  $3+1$ or higher dimensions. 
The analysis is completely parallel to lower dimensional cases, but the physical conclusion is completely different. An odd number of Majorana fermion zero modes arises in \emph{inconsistent} models in $3+1$ or higher dimensions. In chapter \ref{chap:consistency} we generalize this observation: An odd number of Majorana fermion zero modes signifies an inconsistency in $3+1$ or higher dimensions.

\subsection{'t Hooft-Polyakov Monopole in $3+1$ Dimensions}

There is a famous soliton called the 't Hooft-Polyakov monopole in the $3+1$-dimensional $SU(2)$ gauge theory with the Lagrangian density
\begin{align}\label{Lb-SU(2)}
	\cL_b=\frac{1}{4}{\rm tr}F_{\mu\nu}F^{\mu\nu}+\frac{1}{2}{\rm tr}D^{\mu}\Phi D_{\mu}\Phi
\end{align}
where $\Phi$ is a scalar in the adjoint representation of $SU(2)$. 

The 't Hooft-Polyakov monopole solution is
\begin{align}
	\Phi(x)=a\frac{x^iT_i}{r}f(r)
\end{align}
where $T_i$ are the representation matrices of $SU(2)$ and $a$ is a constant. $f(r)$ satisfies
\begin{align}
	f(r)\xlongrightarrow{r\rightarrow 0}0,\,\,\, f(r)\xlongrightarrow{r\rightarrow\infty}1.
\end{align}
$SU(2)$ gauge symmetry is broken to $U(1)$. The unbroken part is
\begin{align}
	F^{U(1)}_{\mu\nu}:={\rm tr}F_{\mu\nu}\Phi/a.
\end{align}

The energy of this configuration is
\begin{align}\label{BPS-bound}
	\int d^3x({\rm tr}B^2/2+{\rm tr}(D\Phi)^2/2)\geq\int d^3x {\rm tr}(B_iD_i\Phi)=\int_{S^2}d\mathbf{S}\cdot {\rm tr}(\mathbf{B}\Phi)=\frac{1}{a}\int_{S^2}d\mathbf{S}\cdot {\rm tr}\mathbf{B}^{U(1)}.
\end{align}
The inequality is saturated if and only if
\begin{align}
	B_i^a=D_i\Phi^a.
\end{align}
This is called the Bogomol'nyi-Parasad-Sommerfield (BPS) equation and the inequality (\ref{BPS-bound}) is called the BPS bound. We assume that the BPS equation is satisfied. This fixes the gauge field configuration and the function $f(r)$. 

The $U(1)$ magnetic charge of this configuration is
\begin{align}
	g=\int_{S^2}d\mathbf{S}\cdot \mathbf{B}^{U(1)}=\frac{1}{a}\int d^3x {\rm tr}(D_i\Phi)^2\propto a.
\end{align}
There are also generalizations of this monopole to arbitrary charges $g\propto na,$ $n\in\bZ$. This $\bZ$ is associated with the topology of the configuration space of the vacua $\pi_2(SU(2)/U(1))=\bZ$. In this sense, the 't Hooft-Polyakov monopole is an analytic realization of the Dirac monopole.



\subsection{Fermion Zero Modes on an 't Hooft-Polyakov Monopole}

Let us add a Dirac fermion $\psi$ in some representation of $SU(2)$ coupling with bosons as
\begin{align}
	\cL=\cL_b+\overline{\psi}\slashed{D}\psi-\overline{\psi}\Phi\psi.
\end{align}
The Dirac equation is
\begin{align}
	(\slashed{D}-\Phi)\psi=0.
\end{align}
The fermion zero modes correspond to the static normalizable solutions of this equation.

In the 't Hooft-Polyakov monopole background of charge $n$, the Callias index theorem \cite{Callias:1977kg} tells us that there are $n$ complex fermion zero modes for a fundamental fermion and $2n$ for an adjoint fermion. See appendix \ref{sec:Callias-index} for the theorem. The quantization of them assigns degenerate local Hilbert space and gives quantum numbers such as spin to the monopole. Let us see the case with a single adjoint fermion in the background $n=1$.

We have two sets of creation/annihilation operators. They originate from the fermion with charges $(\mathbf{2},\mathbf{3})$ under the original $SO(3)$ space rotation and the $SU(2)$ gauge symmetry. Note that the existence of the monopole twists the space rotational group from the original $SO(3)$ to the diagonal subgroup of $SU(2)\times SO(3)$. Since $\mathbf{2}\times\mathbf{3}=\mathbf{2}+\mathbf{4}$, the two zero modes should form $\mathbf{2}$ under the new rotational group. Therefore they carry angular momentum $\pm 1/2$. There are four degenerate monopole ground states
\begin{align}
	\ket{\Omega},a_{1/2}^{\dagger}\ket{\Omega},a_{-1/2}^{\dagger}\ket{\Omega},a_{1/2}^{\dagger}a_{-1/2}^{\dagger}\ket{\Omega},
\end{align}
and they carry spins $0,1/2,-1/2,0$. As a result, fermion zero modes determine the spins of the monopole states and therefore the statistics of them. See section 4 of \cite{Harvey:1996ur} for more details.

\subsection{'t Hooft-Polyakov Monopole and Witten Anomaly}

Instead of a Dirac fermion, we could add a single Weyl fermion $\psi$ in the doublet of $SU(2)$ with the coupling $\overline{\psi}\Phi\psi$. There is a single Majorana fermion zero mode on an 't Hooft-Polyakov monopole of charge $1$, according to the Callias index theorem. 

As in the $2+1$-dimensional case, one might expect that the monopole shows non-abelian statistics. However, the existence of anyons is prohibited in $3+1$ dimensions. Something must be wrong.

This is completely in agreement with the discussion in the section \ref{sec:non-pert-QFT}; This theory has the $SU(2)$ Witten anomaly.  Generally, when we have an odd number of Weyl fermions in the doublet, the theory has the Witten anomaly, and on the other hand, there are an odd number of fermion zero modes on a monopole. The two consistency conditions match well. This relation between the Witten anomaly and an odd number of fermion zero modes was pointed out years ago \cite{McGreevy:2011if}.

This observation implies that an odd number of Majorana fermion zero modes signifies an inconsistency of the theory. In the next chapter \ref{chap:consistency}, we will give a general argument why this is so, under some assumptions.

\subsection{Inconsistent Model in $4+1$ Dimensions}

There is also an inconsistent model in $4+1$ dimensions with an odd number of fermion zero modes on a soliton. Take an $SU(2)$ gauge theory with $n$ fermions in the doublet with the symplectic-Majorana condition in $4+1$ dimension.
When $n$ is odd, this model has the $SU(2)$ Witten anomaly in $4+1$ dimension introduced in \ref{subsec:4+1-non-pert-QFT}.

We now note that an $SU(2)$ instanton configuration on $\bR^4$ can be regarded 
as a point-like soliton in a $4+1$-dimensional theory.
In this background, the index theorem tells us that there are $n$ Majorana fermion zero modes. Therefore, we again find that an odd number of Majorana fermion zero modes leads to an inconsistency of the theory.



\chapter{Majorana Fermion Zero Modes and Consistency}\label{chap:consistency}


We will describe our main new results in this and the next two chapters. In this chapter we will present the generalization of the observations in the previous chapter, namely the general argument that having an odd number of fermion zero modes on a soliton signifies an inconsistency in a theory with $3+1$ or higher dimensions. The important remark is that the argument is not applicable to $2+1$ and $1+1$ dimensions, which allow the existence of anyons. The discussion in this chapter is based on the author's work \cite{Sato:2022vii}

\section*{Setups and Assumptions}

We present an argument that one can never have an odd number of Majorana fermion zero modes on a point-like soliton in a consistent quantum theory in $d+1$ dimension, when $d\ge 3$. We make the following assumptions:
\begin{quote}
	\begin{itemize}
		\item The point-like soliton is dynamical. \emph{Static, external} solitons are removed from the consideration.
		\item The soliton is truly point-like. Namely, it has no additional structures, such as orientational degrees of freedom.
		\item There is the fermion parity symmetry.
	\end{itemize}
\end{quote}
Later we will comment on what happens when these assumptions are removed.

\section*{Proof}

Suppose, on the contrary, that we have a soliton with an odd number, say $n$, of Majorana fermion zero modes, which we denote by $\psi_{1,\ldots,n}$.
It is impossible to assign a local Hilbert space associated with the soliton. 
We can still assign a Hilbert space to a pair of such solitons.
As there are $2n$ Majorana fermion zero modes in total, this Hilbert space has dimension $2^n$. Up to this point, there is no difference whether $d=1,2$ or $d\geq3$.

We now consider semi-classical quantization of this pair of solitons.
When we hold them at a large distance $L$ from each other,
the configuration space has the topology $S^{d-1}/\bZ_2$,
where the quotient comes from our assumption that we consider two identical solitons.
Recall that we assumed $d\ge 3$. 
Then  $\pi_1(S^{d-1}/\bZ_2)=\bZ_2$,
and the wavefunction can be thought of as a wavefunction on $S^{d-1}$ invariant under a $\bZ_2$ operation $P$.
As is well known, this was why there is the boson-fermion dichotomy
when $d\ge 3$,
depending on whether $P$ is realized as $+1$ or $-1$.

Let us first consider the case $n=1$ for simplicity.
The Majorana fermion zero modes  $\psi_{i,j}$ 
supported at each soliton act on $\bC^2$, satisfying $\{\psi_i,\psi_j\}=2\delta_{ij}$.
We then need an order-2 operation $P$ satisfying 
\begin{equation}
P\psi_1 P^{-1}\sim \psi_2,\qquad
P\psi_2 P^{-1}\sim \psi_1 \label{bar}
\end{equation}
and commuting with the fermion parity.
This is clearly impossible. Recall the discussion in \ref{subsec:superconductor-soliton}.
Any such $P$ commuting with the fermion parity $(-1)^F=i\psi_1\psi_2$ must act as
\begin{align}
	(\psi_1,\psi_2)\xrightarrow{\sim}(\psi_2,-\psi_1)
\end{align} 
and is of order 4.
This is against our requirement\footnote{%
Another more geometrical way to state the issue is the following.
We have a bundle of Clifford algebras generated by $\psi^{(1,2)}$ over the configuration space $S^{d-1}/\bZ_2$.
We then ask whether there is a bundle of two-dimensional representations of these Clifford algebras over $S^{d-1}/\bZ_2$.
What we showed here is that there is no such bundle of representations when $d\ge 3$.
} that $P$ is of order 2.

More generally, such $P$ exchanging two sets of $n$ Majorana fermion zero modes is of order 4 when $n$ is odd.
This means that we cannot consistently form the wavefunction of a system of two identical solitons, each having an odd number of Majorana fermion zero modes,
if the spacetime has dimension $3+1$ or higher.

\section*{Comments}

\begin{enumerate}
	\item We cannot apply this argument to $2+1$ and $1+1$ dimensions, because $\pi_1(S^1)=\bZ$ and $\pi_1(S^0)$ is trivial. This is why anyons are allowed in $2+1$ dimensions. In $d=1$ we do not have a concept of spin. 
	
	One might understand more easily the distinction between lower dimensions and higher dimensions in terms of the representation theory, which is essentially the same as our argument. Any massive field should be locally in a representation of infinitesimal rotation algebra $so(d)$, that is, globally in a bundle of a representation of the universal covering\footnote{
		The universal covering of $SO(d)$ is the simply-connected lie group that has the same lie algebra as $SO(d)$, where the simply-connectness means that the fundamental group is trivial. It is the maximal group that has the same lie algebra.
	} of $SO(d)$. When $d\geq3$ the universal covering of $SO(d)$ is $Spin(d)$ which is the double-covering of $SO(d)$. This is why only bosonic and fermionic statistics are allowed in $d\geq3$. The order 4 exchange operation is not allowed. On the contrary the universal covering of $SO(2)$ is $\bR$ which is the infinite-covering of $SO(2)$. It indicates that any statistics are possible in $d=2$.

	\item Note that it is perfectly possible to have an odd number of
	Majorana fermion zero modes on \emph{static, external} point-like solitons.
	Indeed, we can simply consider a field theory where only fermion fields are dynamical,
	while gauge fields and scalars are considered as background fields in the models we considered in section \ref{sec:3+1-MFZ}.
	We can still introduce a monopole background in such a theory,
	which might have an odd number of Majorana fermion zero modes.

	In particular, a point-like object with infinite mass may have an odd number of Majorana fermion zero modes. Such objects are realized by brane intersections for example. A brane stretching infinitely, and therefore with infinite mass, may be a point-like object for the worldvolume theory of the other brane. Such ``point-like'' branes may have an odd number of fermion zero modes. In chapter \ref{chap:branes-intersect} we will see such systems. \label{comment:static-consistency}

	\item A point-like object with an extra structure may have an odd number of Majorana fermion zero modes. In $3+1$ dimensions, Teo and Kane \cite{Teo:2009qv} constructed a point-like hedgehog structure with additional orientational degrees of freedom, on which there is a single Majorana fermion zero mode. Freedman {\it et al.} \cite{Freedman:2010ak} showed that the object has so-called projective ribbon statistics, leading to the consistent exotic statistics in $3+1$ dimensions. 
	
	The existence of such an object is allowed. In the discussion above, we assumed that the configuration space is $S^{d-1}/\bZ_2$, which is not the case for objects with extra structures. Therefore it is out of our scope.\label{comment-structure-consistency}

	\item It is natural to associate this inconsistency with the $\bZ_2$-valued non-perturbative gauge anomaly. In section \ref{sec:3+1-MFZ} we have seen the direct connection between the $\bZ_2$-valued Witten anomaly and an odd number of Majorana fermion zero modes. Therefore, the analysis of fermion zero modes on a point-like soliton is one way to check the cancellation of a class of non-perturbative anomalies in the theory. In the next two chapters, we will apply this criterion to string and M theories.
\end{enumerate}


\chapter{Fermion Zero Modes on Wrapped Branes}\label{chap:branes-wrap}


Let us now focus on the solitons in string and M theory. The solitons in string and M theory are realized by D-branes and M-branes. 

D-branes are extended objects in string theory. D-branes with space dimensions $p\geq-1$ are called D$p$-branes\footnote{
	D($-1$)-branes are understood as instantons.
}. Open strings can end on the D-branes and give the spectrum on the D-branes, including the fermion zero modes on them. M-branes are analogs of D-branes in M theory. There are two kinds of M-branes; M2-branes and M5-branes. Although there is no open string in the description of M theory, M-branes also have spectrums including the fermion zero modes.

Compactifying the branes to $0+1$ dimension, they become point-like. See Figure \ref{fig-branes-wrap}.
Having an odd number of Majorana fermion zero modes on them signifies an inconsistency of the configuration in $3+1$ or higher dimensions. 
This analysis gives a strong tool to check the consistency of string vacua with branes. 
It can also be regarded as one of the criteria for the cancellation of non-perturbative anomalies in string vacua in a simple setting. In this and the next chapters, we will see how this kind of anomaly cancels in string and M theories.
\vspace*{10pt}

\begin{center}
	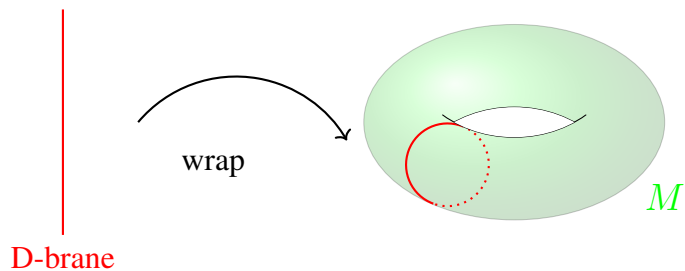
\begin{figure}[h]
		\caption{A D-brane compactified to $0+1$-dimension. A D$p$-brane with $p\geq1$ originally stretches infinitely in the space. When it is wrapped around a cycle in the internal compact manifold $M$, it becomes point-like in the non-compact spacetime.}
		\label{fig-branes-wrap}
		\begin{tikzpicture}[baseline=(current bounding box.center)]
			\draw[red,thick](0,3)--(0,0)node[below]{D-brane};
			\draw[->,thick](1,1.5)arc[radius=1.7,start angle=140,end angle=30];
			\draw(2,1.2)node[below]{wrap};
			\draw[ball color=green,opacity=0.2](6,1.5)circle[x radius=2,y radius=1.3];
			\draw[color=green](8,0.5)node{\large $M$};
			\draw[red,thick](5.3,1.45)arc[radius=0.55,start angle=70,end angle=250];
			\draw[dotted,red,thick](5.3,1.45)arc[radius=0.55,start angle=70,end angle=-110];
			\begin{scope}
				\clip (6,0.6) ellipse (1.5 and 1.3);
				\draw(6,2.6) ellipse (1.5 and 1.3);
				\clip (6,2.6) ellipse (1.5 and 1.3);
				\draw (6,0.4) ellipse (1.5 and 1.3);
				\fill[white] (6,0.4) ellipse (1.5 and 1.3);
			\end{scope}
		\end{tikzpicture}
	\end{figure}
\end{center}

In this chapter, we discuss the fermion zero modes on a single (coincident) M5-brane(s) wrapped around a five-dimensional cycle $Y$. 
We will show that there must be an even number of Majorana fermion zero modes on such a brane because of a particular symmetry of the worldvolume theory in section \ref{sec:R-branes-wrap}. In section \ref{sec:geom-branes-wrap}, we mention the direct calculation of zero modes using the anomaly polynomial of the M5-branes for the case $Y=S^1\times P$, which relies on geometric tools more. This chapter is mainly based on \cite{Sato:2022vii} by the author.

\section{R-symmetry and the Cancellation of the Anomaly}\label{sec:R-branes-wrap}

Let us begin with the situation where a single M5-brane\footnote{
	In this section we restrict our attention to a single M5-brane. The theory on multiple coincident M5-branes is considered to be non-Lagrangian so far, and we do not have weakly-coupled descriptions of it. The geometric argument in the next section is applicable to such coincident M5-branes.
} stretches along $\bR^{1,5}$ in the flat $10+1$-dimensional spacetime $\bR^{1,10}$. The theory on the brane is $5+1$-dimensional $\cN=(0,2)$ supersymmetric theory.
The spectrum of the M5-branes is the $5+1$-dimensional tensor multiplet: a self-dual 2-form field $B$, five real scalar fields $\phi^i$, and two Weyl fermions $\psi^i$. This theory has $SO(5)_R$ R-symmetry corresponding to the normal directions. It acts on $B$ trivially, on $\phi$ as the fundamental representation, and on $\psi$ as the real Dirac representation.

We compactify the total M-theory on $X$ and wrap the M5-brane around the 5-cycle $Y\subset X$. We assume $X$ is a $(10-d)$-dimensional compact manifold. Then the M5-brane is a point-like soliton in $d+1$-dimensional theory. The configuration is inconsistent if there are an odd number of massless fermions on the compactified theory for $d\geq3$, which turns out not to be the case.

The proof that such configuration must have an even number of Majorana fermion zero modes is simple. The $SO(5)_R$ R-symmetry of the theory on M5-branes is broken to $SO(d)_R$ R-symmetry. The massless fermions on the compactified theory must be in the real Dirac representation of this $SO(d)_R$. Therefore, there must be an even number of Majorana-fermions for $d\geq 2$ from the structure of spin representations (See appendix \ref{app:fermions}). Note that this argument kills the possibility of an odd number of fermion zero modes for $d=2$, which is allowed in our general argument in chapter \ref{chap:consistency}.

This argument does not hold for the case $d=1$. We could not kill the possibility for $d=1$, but we have not been able to construct the models with an odd number of Majorana fermion zero modes in $1+1$-dimensional compactifications of M-theory either.

\section{Geometric Calculation using the Anomaly Polynomial}\label{sec:geom-branes-wrap}

In this section, we assume that the compactification manifold is $X=S^1\times M$, and the M5-branes are wrapped around $Y=S^1\times P$, where $P$ is a 4-cycle in $M$. 
Reducing the radius of $S^1$, this situation is equal to type IIA string theory with a (coincident) D4-brane(s) wrapped around the 4-cycle $P$. 
In this case, we can directly calculate the the number of fermion zero modes mod 2. 

The result agrees with the general argument for a single brane in the previous section. We consider it worth mentioning how to calculate the number of fermion zero modes in several ways. Furthermore, the method here also applies to coincident M5-branes, which do not admit weakly-coupled descriptions.

The condition that there cannot be an odd number of Majorana fermion zero modes on a point-like soliton in $d+1$ dimensions
then translates to the condition that there cannot be an odd number of Majorana-Weyl fermions on a string-like soliton in $(d+1)+1$ dimensions.
For simplicity, we assume that the worldsheet theory is almost free 
at some scale along the renormalization group flow, with scalars and fermions.
Then, the number of Majorana-Weyl fermions is even or odd depending on whether
the difference of central charges  $c_L-c_R$ is an integer or an integer plus $1/2$.

Let us begin with the case of a single M5-brane wrapped around a 4-cycle $P$.
Here we adopt the approach of integrating the anomaly polynomial over the 4-cycle. 

The anomaly polynomial of a single M5-brane is \cite{Witten:1996hc,Monnier:2013rpa}
\begin{align}
    \label{6dAnomPoly}
    I_8=\frac{1}{48}\left[p_2(NW)-p_2(TW)+\frac{1}{4}(p_1(TW)-p_1(NW))^2\right]+\frac12\iota^*(G)^2,
\end{align} 
where $W$ is the worldvolume of the M5-brane; $N$ and $T$ are for the normal and tangent bundles, 
$G$ is the background 4-form flux of the spacetime,
and $\iota^*$ denotes the pull-back to the worldvolume.

Let us now integrate it over the 4-cycle $P$.
The 6d worldvolume theory reduces to the 2d worldsheet theory.
Denoting the Chern roots to $TP$ by $\pm\lambda_1$, $\pm\lambda_2$
and those to $NW$ by $\pm n_1$, $\pm n_2$, $0$,
we have
\begin{align}
    I_8=\frac{1}{48}\left[n_1^2n_2^2-\lambda_1^2\lambda_2^2-(\lambda_1^2+\lambda_2^2)p_1(T\Sigma)+\frac{1}{4}(\lambda_1^2+\lambda_2^2+p_1(T\Sigma)-n_1^2-n_2^2)^2\right]+\frac12\iota^*(G)^2.
\end{align}
Here, $T\Sigma$ is the tangent bundle of the worldsheet. Therefore, the anomaly polynomial of the worldsheet theory is given by 
\begin{align}
    I_4
    &=-\frac{p_1(T\Sigma)}{96}\int_P\left[\lambda_1^2+\lambda_2^2+n_1^2+n_2^2\right].
\end{align}
Here $G^2$ did not contribute, since we assume that the flux is only along $M$.

Recalling that $c_L-c_R$ and the anomaly polynomial of a worldsheet theory are related as \begin{equation}
I_4 = \frac{c_L-c_R}{24}p_1(T\Sigma)
\end{equation} in general, we see that \begin{align}
    \label{cGB}
    c_L-c_R=-\frac{1}{4}\int_P(\lambda_1^2+\lambda_2^2+n_1^2+n_2^2)=-\frac{1}{4}\int_P\iota^*p_1(TM).
\end{align}
When $M$ is a Calabi-Yau 3-fold, 
$c_L-c_R=\frac{1}{2}\int_P\iota^*c_2(TM)$ for an M5-brane because of the relation $p_1=c_1^2-2c_2$ on a complex manifold and the Calabi-Yau condition 
$c_1(TM)=0$. 
This can then be compared with the $c_L$, $c_R$ found in \cite{Witten:1996hc} and we find a nice agreement.

Our question is whether this $c_L-c_R$ given in \eqref{cGB} is an integer. 
For a spin manifold $M$, $p_1(TM)/2$ is known to be an integral class\footnote{%
This is because $p_1=w_2^2$ mod 2 for any orthogonal bundles,
and $w_2$ of a spin bundle is zero.
}.
Therefore, $c_L-c_R$ given in \eqref{cGB} is at worst a half-integer.
To see that this is actually an integer, we use two facts about M-theory.
The first is the shifted quantization law of the $G$-flux, 
originally found in \cite{Witten:1996md}: \begin{equation}
[G] - [\frac{p_1(TM)}4] \in H^4(TM,\bZ)
\end{equation}  
and the second is the Bianchi identity \cite{Howe:1996yn,Howe:1997fb,Chu:1997iw}
\begin{equation}
 -dH=\iota^*G
\end{equation} 
on the worldvolume of the M5-brane, where $H$ is the flux of the self-dual 2-form defined globally on it.
These two conditions mean that \begin{equation}
[\frac{\iota^*p_1(TM)}4] \in H^4(TP,\bZ), 
\end{equation}
which was what we wanted to show.\footnote{%
We note that when $M$ is simply-connected, spin and has no torsion in cohomology, 
Wall's theorem \cite{Wall}
says that $\int_M(4P^3-P\cdot p_1(TM))\in 24\bZ$.
This then means that $p_1(TM)/4$ is an integral class when pulled back to $P$.
The argument in the main text is applicable  more generally, in that $M$ may not be necessarily simply-connected,  may have torsion, and may be of any dimensions.
}

Next, let us analyze the situation where coincident M5-branes are wrapped around the 4-cycle $P$. The anomaly polynomial for the M5-branes of type $H$\footnote{
	$H$ is either of $A_n$, $D_n$, $E_n$.
} is \cite{Harvey:1998bx,Intriligator:2000eq,Yi:2001bz}
\begin{align}
    I_8(G)=r_HI_8+d_Hh_H\frac{p_2(NW)}{24}
\end{align}
where $r_H,d_H,h_H$ are the rank, the dimension, and the Coxeter number of $H$. See Table \ref{numbers-H}.

The calculation is the same as before, and we get
\begin{align}
    c_L-c_R=-\frac{r_H}{4}\int_P(\lambda_1^2+\lambda_2^2+n_1^2+n_2^2)=-\frac{r_H}{4}\int_P\iota^*p_1(TM).
\end{align}
This is an integer because of the same reason as before.

\begin{table}[h]
	\caption{$r_H$, $d_H$, $h_H$ for various Lie groups.}\label{numbers-H}
    \centering
    \begin{tabular}{|c|c|c|c|}
        \hline
        $H$&rank $r_H$&dimension $d_H$&Coxeter number $h_H$\\
        \hline \hline
        $A_n$ ($SU(n+1)$)&$n$&$n(n+2)$&$n+1$\\
		$D_n$ ($SO(n)$)&$n$&$n(2n-1)$&$2n-2$\\
		$E_6$&6&78&12\\
		$E_7$&7&133&18\\
		$E_8$&8&248&30\\
        \hline
    \end{tabular}
\end{table}

\subsection*{Remarks}
\begin{itemize}
	\item Recall the discussion in the previous section \ref{sec:R-branes-wrap}. Combined with the calculation in this section, it is natural to speculate that M5-branes wrapped around 5-cycles always have an even number of Majorana fermion zero modes, with a little exception in the case $d=1$. 
	\item Our analysis in this section can be thought of as providing another piece of evidence that the Witten anomaly does not arise in the type IIA string and M theories,
	where the $SU(2)$ gauge group arises from the $\bC^2/\bZ_2$ singularity.
	In such cases, the spontaneous breaking of symmetry to $U(1)$ would be given by a resolution of the singularity, making the internal manifold smooth.
	Then the 't Hooft-Polyakov monopole would be given by a wrapped D4-brane, on which we showed that there are an even number of fermion zero modes.
\end{itemize}




\newpage


\chapter{Fermion Zero Modes at Brane Intersections}\label{chap:branes-intersect}


In this chapter, we analyze the configurations where multiple D-branes and orientifold planes intersect in the spacetime. 
The intersections of branes have additional fermion zero modes coming from open strings stretching between them. See Figure \ref{fig-branes-intersect}.
For orientifold-planes, see the setups below in section \ref{sec:setup-branes-intersect} and appendix \ref{sec:orientifold-spectrum}.
We see some configurations have an odd number of Majorana fermion zero modes on a point-like intersection. See appendix \ref{app:open-spectrum} for the detailed calculations of branes. This chapter is mainly based on \cite{ST:2024}.

In section \ref{sec:setup-branes-intersect}, we introduce setups of our interest and state our criterion on the consistency of brane intersections. 
In sections \ref{sec:ND=8-branes-intersect}, \ref{sec:ND=6-branes-intersect} and \ref{sec:ND=4-branes-intersect}, we analyze various configurations of intersecting branes.
In section \ref{sec:interpret-brane-intersect}, we give another interpretation of the inconsistency of having an odd number of Majorana fermion zero modes at brane intersections in the language of  branes. Combined with the analysis in the previous chapter \ref{chap:branes-wrap}, it suggests the consistency conditions of string theories already prohibit the possibilities of having an odd number of Majorana fermion zero modes on a point-like brane.
In section \ref{sec:field-brane-intersect}, we make a little digression and describe the calculation method of fermion zero modes in the language of field theory on branes.

It is also meaningful to investigate the possibility of having an odd number of Majorana fermion zero modes in $2+1$ dimensional compactifications of string theories, which leads to the realization of non-abelian anyons in string theories. To the author's knowledge, the direct realization of anyons in string theories is not known so far. We give some discussions and comments on the problem at the end of section \ref{sec:interpret-brane-intersect}.

\begin{center}
	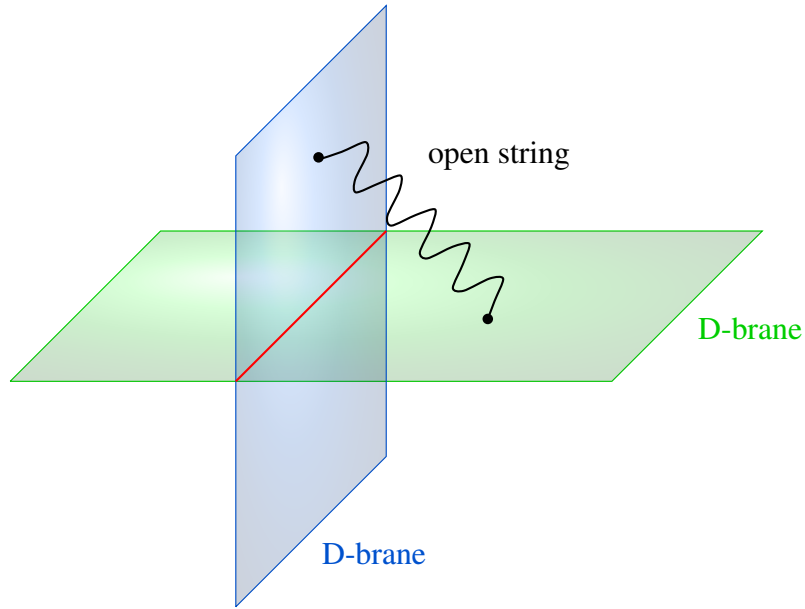
\begin{figure}[h]
		\caption{An open string stretching between two D-branes. The two D-branes intersect along the red line.}
		\label{fig-branes-intersect}
		\begin{tikzpicture}[baseline=(current bounding box.center)]
			\draw[green](0,0)--(8,0)--node[below right]{D-brane}(10,2)--(2,2)--(0,0);
			\draw[ball color=green,opacity=0.2](0,0)--(8,0)--node[below right]{D-brane}(10,2)--(2,2)--(0,0);
			\draw[blue](3,-3)--node[below right]{D-brane}(5,-1)--(5,5)--(3,3)--(3,-3);
			\draw[ball color=blue,opacity=0.2](3,-3)--node[below right]{D-brane}(5,-1)--(5,5)--(3,3)--(3,-3);
			\draw[thick,red](3,0)--(5,2);
			\draw[thick,domain=3:6.1,smooth]plot[rotate around={-45:(5,1)}](\x,{2.07+0.3*sin(10*\x r)});
			\draw(6.5,3)node{open string};
			\fill[black](4.1,2.98)circle[radius=0.07];
			\fill[black](6.35,0.83)circle[radius=0.07];
		\end{tikzpicture}
	\end{figure}
\end{center}

\section{Setups and Criterion}\label{sec:setup-branes-intersect}

The typical setup is as follows. There are two D-branes with spatial dimensions $p,q$ \footnote{
	We call them D$p$-brane and D$q$-brane in the general discussions, or when the dimensions of them are different. If $p=q$, we distinguish them by the subscripts as D$p_1$ and D$p_2$.
}
stretching infinitely in the flat spacetime of type II string theory. We assume that the two branes coincide along some directions, and intersect vertically in other directions. 

The boundary conditions of the oscillators on open strings stretching between the two D-branes have three possibilities; NN, ND, and DD boundary conditions, where N and D stand for the Neumann and Dirichlet boundary conditions at an end of the open string. We can identify the D-branes configuration by identifying $\#$NN, $\#$ND, $\#$DD. Obviously they satisfy $\#{\rm NN}+\#{\rm ND}+\#{\rm DD}=10$, and $\#$NN$\geq1$ because the time direction is NN. 

We also take an orientifold projection, namely, we identify states under a combined transformation of worldsheet parity transformation and a spacetime reflection. The fixed point set of the reflection is called an orientifold-plane. Orientifold-planes with space dimensions $r$ are called O$r$-planes. There are two types of orientifold projections; O$r^+$-planes and O$r^-$-planes. On coincident $N$ D$p$-branes contained in a O$r^+$(O$r^-$)-plane, $Sp(N/2)$ ($SO(N)$) gauge symmetry arises when $p=r$ and $SO(N)$ ($Sp(N/2)$) gauge symmetry arises when $p=r-4$. See appendix \ref{sec:orientifold-spectrum} for more details.

We assume that D$p$ and D$q$-branes are both contained in the orientifold plane. The coexistence of a single D$p$-brane and an O$r^-$(O$r^+$)-plane is only possible when $p-r=8n(+4)$, where the orientifold mirror of the D$p$-brane is itself. Sometimes such a configuration is called a half D$p$-brane, because the RR-charge of it is 1/2 of the ordinary brane. We focus on the intersections of two half D-branes as the candidates of an odd number of fermion zero modes\footnote{
	When $p-r\neq8n(+4)$, the D$p$-brane must make a pair with another D$p$-brane or an anti D$p$-brane to coexist with the O$r$-plane. The existence of another (anti) brane doubles the fermion zero modes. Therefore, to reduce the degrees of freedom on an open string, half D$p$-branes are the most efficient objects.
}.

Let us focus on the theory on the D$p$-brane. D$q$-brane can be thought of as an extended object with dimension $\#$NN$-1$ in the worldvolume theory on D$p$-brane. The massless fermionic degrees of freedom on the D$q$-brane come from open strings ending on the D$q$-brane on both sides, and open strings stretching between D$p$ and D$q$-brane. By counting them, we can calculate the number of fermion zero modes on the D$q$-brane\footnote{
	Sometimes we consider the intersection of coinciding branes, $N$ coinciding D$p$-branes and $M$ coinciding D$q$-branes. The fermion zero modes on the open string stretching between them form bifundamental representations of $G_p\times G_q$, where $G_{p/q}$ are the gauge symmetry of the worldvolume theory on D$p/q$-branes.
}. According to the calculation in the appendix \ref{app:open-spectrum}, the number of fermion zero modes depends only on $\#$ND. Therefore taking T-dual does not change the number of fermion zero modes. We will ignore the contributions from open strings ending on D$q$-brane on both sides because they are sufficiently even.

Suppose there are an odd number of Majorana fermion zero modes on D$q$-brane and the worldvolume theory on D$p$-brane is $3+1$ or higher dimensional. When $q\geq1$, the situation is totally acceptable because the brane stretches infinitely and therefore has infinite mass. 
On the other hand, when $q=0$ the configuration is inconsistent according to the argument in chapter \ref{chap:consistency}. 
One might wonder if this contradicts the fact that taking T-dual changes $q$ without changing the overall physics including fermion zero modes\footnote{
	The similar situation can be found in \cite{Hyakutake:2000mr}, which discusses the existence of coinciding D$p$-O$p^-$ bound states. The bound states are allowed for $p\leq6$ whereas not allowed for $p\geq7$. In that case, the consistency condition of Wilson lines on the compactified dimension determines whether such compactification is allowed or not.
}.  The resolution of this puzzle lies in the necessity to compactify the direction to $S^1$ to take T-dual.

\emph{Here comes our criterion}: An odd number of Majorana fermion zero modes on a D$q$-brane stretching infinitely in the spacetime is possible for $q\geq1$, but compactifying the brane to $0+1$ dimension leads to an inconsistency. Therefore such compactifications are not allowed.

Let us see some examples for various $\#$ND\footnote{
	We comment on the relations of our discussion to the work \cite{Ryu:2010fe}, trying to realize topological insulators by multiple branes. They also calculated fermion zero modes to clarify the corresponding type of topological insulators. The situations they consider are different from our situations in that D-branes are separated and therefore the lightest fermions have mass proportional to the distance. They also realize massless fermions by bending the branes. The realizations of fermion zero modes by bending branes are not our interests from two points of view: They are infinitely massive in $2+1$ or higher dimensions, and they have the additional structure of winding. See the comment \ref{comment-structure-consistency} in chapter \ref{chap:consistency}. Nevertheless, they motivated our investigation of intersecting D-branes. Some of our examples are from their work.}
. Our focus is the cases $\#\text{ND}=8,6,4$.

\section{$\#$ND $=8$}\label{sec:ND=8-branes-intersect}

According to the appendix \ref{app:open-spectrum}, the massless spectrum of an open string stretching between two intersecting D-branes is one single Majorana fermion in $0+1$-dimensional basis for each orientation of the open string when $\#$ND $=8$. Therefore, by adding some orientifold O$r$, we immediately get an odd number of fermion zero modes on a brane. 

The dimension of D-branes must be $r+8n(r+8n+4)$, $n\in\bZ$, so that a single D-brane can coexist with O$r^-$(O$r^+$)-plane. Only two configurations up to T-dual satisfy this condition: D1-D9-O9$^-$ and D5-D5-O9$^+$.

\subsection*{D1-D9-O9$^-$}

We begin with the system where a D1-brane and a D9-brane with an O9$^-$-plane stretch in the $9+1$-dimensional spacetime in type IIB theory. See Table \ref{D1-D9-O9} for the configuration, where $\circ$ means that the brane stretches infinitely along the direction. 

As explained above, there are an odd number of Majorana fermion zero modes on the D1-brane. Compactifying D1 into a small circle, we get $8+1$-dimensional worldvolume theory on the D9-brane with the point-like D1-brane. Our argument in the chapter \ref{chap:consistency} tells us that this configuration is inconsistent.

We already know that this configuration is inconsistent because this is the type I $SO(1)$ superstring with a D1-brane. Type I theory must have $SO(32)$ gauge symmetry because of the tadpole anomaly cancellation. So this example is inconsistent even before the compactification, and even without D1-brane. This analysis could be seen as a weaker paraphrase of the tadpole anomaly.

\begin{table}[h]
	\caption{D1-D9-O9$^-$}\label{D1-D9-O9}
    \centering
    \begin{tabular}{|c|cccccccccc|}
        \hline
        directions&0&1&2&3&4&5&6&7&8&9\\
        \hline \hline
        D1&$\circ$&$\circ$&&&&&&&&\\
        D9&$\circ$&$\circ$&$\circ$&$\circ$&$\circ$&$\circ$&$\circ$&$\circ$&$\circ$&$\circ$\\
		O9$^-$&$\circ$&$\circ$&$\circ$&$\circ$&$\circ$&$\circ$&$\circ$&$\circ$&$\circ$&$\circ$\\
        \hline
    \end{tabular}
\end{table}

\subsection*{D0-D8-O8$^-$}

Next is the configuration where a single D0-brane is on coincident D8-brane and O8$^-$-plane. This is the T-dual of the above. This configuration is immediately inconsistent because of the odd number of fermion zero modes on a D0-brane, which is point-like. The coincident D8-O8$^+$ bound state is already prohibited by the discussion in \cite{Hyakutake:2000mr}. Again the configuration is inconsistent without D0-brane.

Note that it is completely acceptable when an even number of D8-branes coincides with an O8$^+$-plane. The number of fermion zero modes on a single D0-brane also becomes even in this case.

\subsection*{D5-D5-O9$^+$}

Next we discuss the situation where $N$ coinciding D5$_1$-branes stretch infinitely along 1-5 direction, and the other $M$ coinciding D5$_2$-branes along 5-9 direction, on a flat O9$^+$-plane. See Table \ref{D5-D5-O9}. The 2 branes intersect along 5-direction. It is known that there must be 32 anti-D9-branes on the O9$^+$-plane so that the anomaly polynomials of D5-branes vanish. Nevertheless, we focus on only D5-branes and O9$^+$-plane because the existence of 32 D9-branes does not affect the even/oddness of the number of the fermion zero modes.





The massless fermionic spectrum on an open string stretching between D5$_1$ and D5$_2$ is a single Majorana fermion in the bifundamental representation of $SO(N)\times SO(M)$. For odd $N$ and $M$, compactifying 5-9 directions, the number of Majorana fermion zero modes on the point-like D5$_2$-branes is odd, which leads to the inconsistency of the configuration.

\subsection*{Remarks}
\begin{itemize}
    \item The configuration is not necessarily inconsistent if we do not compactify all of 5-9 directions. Because both D-branes are infinitely heavy, we do not care about the quantization of them, although D5$_2$-branes are ``point-like'' for the worldvolume theory of D5$_1$-branes after compactifying only the 5-direction. It was one of the exceptions of our argument discussed in the comment \ref{comment:static-consistency} of chapter \ref{chap:consistency}. 
    \item Also, the configuration is consistent after the compactification when $N$ is even. One might think it is also consistent when $M$ is even because the number of fermion zero modes is even. However, it will turn out that the consistency after the compactification only depends on $N$. $N$ has to be even so that the configuration is consistent after the compactification along 5-9 directions. See section \ref{sec:interpret-brane-intersect} for the discussion. 
    \item Taking T-dual along the 5-direction of the above, we get the D4-D4-O8$^+$ system. Note that 32 D9-branes are not needed in this case. This configuration is also considered in \cite{Ryu:2010fe}, realizing the Class AI topological insulator, where the branes were not compactified.
\end{itemize}


\begin{table}[h]
	\caption{D5-D5-O9$^+$}\label{D5-D5-O9}
    \centering
    \begin{tabular}{|c|cccccccccc|}
        \hline
        directions&0&1&2&3&4&5&6&7&8&9\\
        \hline \hline
        $N$ D5$_1$&$\circ$&$\circ$&$\circ$&$\circ$&$\circ$&$\circ$&&&&\\
        $M$ D5$_2$&$\circ$&&&&&$\circ$&$\circ$&$\circ$&$\circ$&$\circ$\\
		O9$^+$&$\circ$&$\circ$&$\circ$&$\circ$&$\circ$&$\circ$&$\circ$&$\circ$&$\circ$&$\circ$\\
        \hline
    \end{tabular}
\end{table}

\subsection*{D8-D0-O4$^+$}

Compactifying along 5-9 directions on $T^5$ and taking T-dual of the above gives the D8-D0-O4$^+$ system (Table \ref{D0-D8-O4}). We comment on this configuration, although this is not included in our setups because D8-branes are not contained in the orientifold plane. This configuration is immediately inconsistent when $NM$ is odd. This observation obviously agrees with the observation above. 

\begin{table}[h]
	\caption{D8-D0-O4$^+$}\label{D0-D8-O4}
    \centering
    \begin{tabular}{|c|cccccccccc|}
        \hline
        directions&0&1&2&3&4&5&6&7&8&9\\
        \hline \hline
        $N$ D8&$\circ$&$\circ$&$\circ$&$\circ$&$\circ$&&$\circ$&$\circ$&$\circ$&$\circ$\\
		$M$ D0&$\circ$&&&&&&&&&\\
		O4$^+$&$\circ$&$\circ$&$\circ$&$\circ$&$\circ$&&&&&\\
        \hline
    \end{tabular}
\end{table}

\section{$\#$ND $=6$}\label{sec:ND=6-branes-intersect}

When $\#$ND $=6$ the massless spectrum of an open string stretching between two D-branes is two Majorana fermions in $0+1$-dimensional basis for each orientation of the string. They form a doublet of Spin(4) corresponding to the NN and DD directions. To get an odd number of fermion zero modes on a brane, we need to take some orbifold and halve the degrees of freedom in addition to the orientifold projection.

Let us recall the ordinary orbifolding procedure. When a theory has $SO(n)$ rotational symmetry, taking the orbifold means gauging the finite subgroup of $SO(n)$. We identify the fields under the transformation
\begin{align}
	\theta:={\rm exp}[2\pi i(\sum \theta_iJ_i)]
\end{align}
for rational numbers $\theta_i$, where $J_i$ are the generators of $so(n)$. When a field $\psi$ is a spinor representation of $SO(n)$, $J_i$ becomes $S_i$. The identification under $\theta$ means the states satisfying
\begin{align}\label{orbifold}
	\sum\theta_is_i\in\bZ
\end{align}
survive under the orbifolding.

\subsection*{D3-D3-O7$^+$}

The simplest example is the D3-D3-O7$^+$ system (Table \ref{D3-D3-O7}). This configuration, up to T-dual, is the only example of the intersecting half D-branes with $\#$ND $=6$. The worldvolume theory on $N$ coinciding D3$_1$-branes is a $3+1$-dimensional $SO(N)$ gauge theory. As discussed above, the fermion zero modes on the coincident $M$ D3$_2$-branes are two Majorana fermions in the bifundamental representation of $SO(N)\times SO(M)$.

The worldsheet CFT on the open string has $SO(3)$ symmetry corresponding to the rotation of DD (7-9) directions. The two fermion zero modes form a doublet under this rotation. Therefore one might think taking orbifold projection halves the degrees of freedom, and ends up with an odd number of fermion zero modes. However, it is clear that orbifolding this system by this $SO(3)$ symmetry will kill all fermion zero modes. $SO(3)$ has only one generator $s_{89}$, and it is impossible to satisfy the condition (\ref{orbifold}) for a single generator (or the condition is trivial). Therefore, we conclude that string theory already prohibits an odd number of Majorana fermion zero modes realized by intersecting half D-branes with $\#$ND $=6$. It is still possible that there is some method to take a nice orbifold of this system, but we have not been able to find such an example yet.



\begin{table}[h]
	\caption{D3-D3-O7$^+$}\label{D3-D3-O7}
    \centering
    \begin{tabular}{|c|cccccccccc|}
        \hline
        directions&0&1&2&3&4&5&6&7&8&9\\
        \hline \hline
        $N$ D3$_1$&$\circ$&$\circ$&$\circ$&$\circ$&&&&&&\\
		$M$ D3$_2$&$\circ$&&&&$\circ$&$\circ$&$\circ$&&&\\
		O7$^+$&$\circ$&$\circ$&$\circ$&$\circ$&$\circ$&$\circ$&$\circ$&$\circ$&&\\
        \hline
    \end{tabular}
\end{table}

\section{$\#$ND $=4$}\label{sec:ND=4-branes-intersect}

Finally we investigate the case $\#$ND $=4$. We have four Majorana fermion zero modes for each orientation.

\subsection*{D2-D2-O6$^+$}

Again there is only one example up to T-dual: D2-D2-O6$^+$ (Table \ref{D2-D2-O6}). There are four Majorana fermion zero modes in the bifundamental representation of $SO(N)\times SO(M)$.

This system has $SO(5)$ symmetry corresponding to the rotation of 5-9 directions. Four fermion zero modes form a spinor representation under the rotation $SO(5)$. We can halve the degrees of freedom by orbifolding as
\begin{align}
	\theta={\rm exp}[i\pi(J_{56}+J_{78})]
\end{align}
which restrict the fermionic spectrum on $s_{56}+s_{78}=0$. The subgroup generated by $\theta$ is the $\pi$ rotation of 5-8 directions. This halves the number of fermion zero modes. Further orbifold projection is not possible because of the same reason as the case D3-D3-O7$^+$. We conclude that an odd number of Majorana fermion zero modes are prohibited also in this case.

\begin{table}[h]
	\caption{D2-D2-O6$^+$}\label{D2-D2-O6}
    \centering
    \begin{tabular}{|c|cccccccccc|}
        \hline
        directions&0&1&2&3&4&5&6&7&8&9\\
        \hline \hline
        $N$ D2$_1$&$\circ$&$\circ$&$\circ$&&&&&&&\\
		$M$ D2$_2$&$\circ$&&&$\circ$&$\circ$&&&&&\\
		O6$^+$&$\circ$&$\circ$&$\circ$&$\circ$&$\circ$&$\circ$&$\circ$&&&\\
        \hline
    \end{tabular}
\end{table}

\section{A New Interpretation of the Inconsistency}\label{sec:interpret-brane-intersect}

We have seen that configurations with an odd number of fermion zero modes are possible for $\#$ND $=8$, whereas impossible for $\#$ND $=6,4$. Let us focus on the specific example D5-D5-O9$^+$, which has a single Majorana fermion zero mode. Recall that the configuration is acceptable when two D-branes stretch infinitely, whereas compactifying one of the branes on $T^5$ and making it into $0+1$-dimensional object leads to an inconsistency of the configuration when $NM$ is odd. 
One natural question is whether such inconsistency can be understood from the original consistency conditions of branes and string theories.

The answer is yes.
There is another interpretation of the inconsistency of having an odd number of fermion zero modes on intersecting D-branes, in terms of the gauge invariance on the branes. 

We consider the situation where $N$ coincident D$p$-branes and $M$ coincident D$q$-branes intersect vertically with some orientifold plane in the spacetime. We assume that gauge symmetries of both D$p$ and D$q$-branes are of type $SO$ as in the previous sections. 

Then suppose there are an odd number (say $n$) of Majorana fermion zero modes on an open string stretching between the two sets of D-branes. Compactification of D$q$-branes on $T^q$ gives a theory on the D$p$-branes with an orientifold and point-like D$q$-branes.
The worldvolume theory on D$p$-branes is $SO(N)$ gauge theory, and there is $SO(M)$ gauge symmetry acting on the fields on D$q$-branes. The fermion zero modes from the open string stretching between them form bifundamental multiplets of the $SO(N)\times SO(M)$. In total there are $nNM$ Majorana fermion zero modes.

The ground states of this configuration should be a spinor representation of the fermion zero modes coming from the open string. To construct ground states, we need to make pairs of fermions $\psi^{a\pm}:=\psi^a\pm\psi^{a+1}$, where $a$ labels the pairs of fermions, and define $\ket{0}$ so that
\begin{align}\label{gs}
    \psi^{a-}\ket{0}=0
\end{align}
for all $a$. 
The ground states are of the form $\prod\psi^{a+}\ket{0}$. Since $SO(M)$ is a gauge symmetry\footnote{
    The condition (\ref{gs}) does not have to be invariant under $SO(N)$. It can be explained as follows. In the Standard Model there is $U(1)\times SU(2)\times SU(3)$ gauge symmetry. We allow the existence of a single electron, which is not invariant under $SU(2)$. A state itself is allowed not to be gauge invariant, whereas physical quantities have to. 
	
	The $SO(M)$ is not a gauge symmetry of the spacetime theory but of the point-like soliton. Therefore, $SO(M)$ should not change the states of the spacetime.
}, (\ref{gs}) should have invariant meaning under $SO(M)$, and $SO(M)$-invariant ground states are physical.

Let us check what the $SO(M)$-invariance implies. We set $n=1$ for simplicity\footnote{
    When $n=2m+1$, we can make $m$ pairs and they can be ignored in this argument.
}. We use superscripts as the index of $SO(N)$, subscripts as $SO(M)$, and denote fermion zero modes as $\psi^n_m$. When $N$ is even, the condition (\ref{gs}) can have an invariant meaning by making pairs $\psi_m^{a\pm}=\psi^a_m\pm\psi^{a+1}_m$. However, when $N$ is odd, we cannot make such pairs. 
As a result, we cannot make consistent ground states, and it is an inconsistency of the theory\footnote{
    When $M=1$, we cannot apply this argument because $SO(1)$ is trivial and therefore all states are invariant. However, if such configurations are allowed, three of them will also be allowed and it leads to the inconsistency.
}. We note that compactification is necessary also in this argument. Without the compactification there is no reason for imposing $SO(M)$ invariance.

We conclude that the number of Majorana fermion zero modes on point-like D-branes intersecting with other branes must always be even because of the gauge invariance of the theory on the branes.
The discussion of gauge invariance on the point-like soliton does not appear in field theories, so this condition is characteristic of string theories. 
Combined with the analysis in the previous chapter \ref{chap:branes-wrap}, this observation implies that the consistency conditions of string theory already prohibit the odd number of Majorana fermion zero modes on a point-like soliton. 

\subsection*{Remarks}
\begin{itemize}
	\item We again remark that it is completely possible that D-branes stretching infinitely have an odd number of Majorana fermion zero modes, also in the argument of the gauge invariance on the branes. Without the compactification there is no reason to impose the gauge invariance on the ground states. It agrees with the criterion we introduced in section \ref{sec:setup-branes-intersect}, and the argument in chapter \ref{chap:consistency}.

    \item Note that this argument depends only on $N$, the number of uncompactified branes. Therefore, even if the number of Majorana fermion zero modes, which is proportional to $NM$, is even, the configuration is inconsistent if $N$ is odd.
	
	\item This argument holds only when there are states in the fundamental representation of the gauge group on the branes. Therefore, we cannot apply the same argument on non-intersecting branes, on which only adjoint fields exist.
	
	\item Combined with the analysis in chapter \ref{chap:branes-wrap}, this observation implies that the consistency conditions of string theory already prohibit having an odd number of Majorana fermion zero modes on point-like branes, at least in all cases we considered. Note that our analysis does not depend on the dimension of the spacetime, except in the case of a single M5-brane with $d=1$. 
	
	\item Therefore, we conclude that anyons of our interest in $2+1$-dimensional systems cannot be realized by the single (coincident) brane(s) or the intersections of branes. This may not imply that the realization of anyons in string theories is impossible, but it should be more complicated if possible.
	
	In section \ref{sec:R-branes-wrap}, we could not kill the possibility of an odd number of Majorana fermion zero modes on a point-like M5-brane in $1+1$ dimension. If such a state is realizable, it will be an analog of Kitaev's quantum wire in M-theory.
\end{itemize}

\section{Field Theory Calculation}\label{sec:field-brane-intersect}

Before we conclude this thesis, let us make a little digression and calculate the number of fermion zero modes by the field theory argument. 
It gives another confirmation of the results above and our calculation methods of open string spectrums. The readers may skip this digression and move to the conclusion \ref{chap:conclusion}.

We focus on the D5-D5-O9$^+$ system with $N=M=1$. First let us start with two coincident D5-branes stretching infinitely along 1-5 directions, with O9$^+$-plane and 32 anti D9-branes filling the spacetime. 
We ignore 32 anti D9-branes because they do not contribute to the even/oddness of the number of Majorana fermion zero modes.

The worldvolume theory of two coincident D-branes is $5+1$-dimensional $SO(2)_G$ gauge theory with $\cN=(0,1)$ supersymmetry. The spectrum is a gauge field $A$, an adjoint Weyl fermion (gaugino) $\Psi$, and four real bosons $\phi^i$ and a Weyl fermion $\psi$ in the symmetric representation of SO(2). This theory has $SO(4)\simeq SU(2)_L\times SU(2)_R$ symmetry corresponding to the rotation of 6-9 directions. See Table \ref{D5-O9-spectrum} for the charges of each field. 

Translating $SO(2)_G$ into $U(1)$, the adjoint representation of $SO(2)$ becomes trivial representation of $U(1)$, and the symmetric representation of $SO(2)$ is decomposed into charge $0,\pm2$ representations of $U(1)$. Consequently, the worldvolume theory of two D5-branes with O9$^+$-plane is a $U(1)$ gauge theory with a neutral Weyl fermion $\Psi$ (gaugino) and four series of real bosons $\phi^i_0,\phi^i_+,\phi^i_-$ and Weyl fermions $\psi_0,\psi_+,\psi_-$ with U(1)-charges 0,2,-2. This theory has Yukawa coupling $\overline{\Psi}\phi^i_+\Gamma^i\psi_--\overline{\Psi}\phi^i_-\Gamma^i\psi_++cc.$ where $i=6,7,8,9$\footnote{
	In the section \ref{sec:example-spectrum} of appendix \ref{app:open-spectrum} we show the detailed calculation for the $2+1$-dimensional case. The calculation here is essentially the same.
}.

\begin{table}[h]
	\caption{Summary of Charges for D5-O9$^+$}
	\label{D5-O9-spectrum}
    \centering
    \begin{tabular}{|c|c|c|c|c|}
        \hline
        &$SO(1,5)$&$SU(2)_L$&$SU(2)_R$&$SO(2)_G$ \\
        \hline \hline
        $A$&$\mathbf{6}_V$&$\mathbf{1}$&$\mathbf{1}$&adj.\\
        $\Psi$&$\mathbf{8}_W$&$\mathbf{1}$&$\mathbf{2}'$&adj.\\
        $\phi$&$\mathbf{1}$&$\mathbf{2}$&$\mathbf{2}'$&sym.\\
        $\psi$&$\mathbf{8}'_W$&$\mathbf{2}$&$\mathbf{1}$&sym.\\
        \hline
    \end{tabular}
\end{table}

To realize the configuration D5-D5-O9$^+$ in the previous section from this setup, we add vacuum expectation values to scalars $\phi^i$ as
\begin{align}\label{mass-term}
    \langle\phi^i(x)\rangle=\begin{pmatrix}
		\alpha x^{i-4}&0\\
		0&0
	\end{pmatrix}
\end{align}
for $i=6,7,8,9$. In the D-brane picture, this means that we rotate one of the D5-branes, say D5$_2$-brane, fixing 5-direction by the degree arctan$\alpha$. See figure \ref{fig-rotate-branes}. Taking $\alpha=\infty$ we get the configuration where D5$_1$-brane is along 1-5 directions and D5$_2$-brane along 5-9 directions. D5$_2$-brane is a string-like object in the field theory on D5$_1$-brane.

\begin{center}
	\begin{figure}[h]
		\caption{Rotate D5$_2$-brane by the degree arctan$\alpha$.}
		\label{fig-rotate-branes}
		\begin{tikzpicture}[baseline=(current bounding box.center)]
			\draw[green](0,0)--(8,0)--node[below right]{D5$_1$-brane}(10,2)--(2,2)--cycle;
			\draw[ball color=green,opacity=0.2](0,0)--(8,0)--(10,2)--(2,2)--(0,0);
			\draw[blue](0.5,-1.5)--node[below right]{D5$_2$-brane}(7.5,1.5)--(9.5,3.5)--(2.5,0.5)--cycle;
			\draw[ball color=blue,opacity=0.2](0.5,-1.5)--(7.5,1.5)--(9.5,3.5)--(2.5,0.5)--cycle;
			\draw[black](4,0)--(6,2);
			\draw[thick,->](8,1)arc[radius=2,start angle=0,end angle=35];
			\draw(10,2.3)node{rotate by arctan$\alpha$};
		\end{tikzpicture}
	\end{figure}
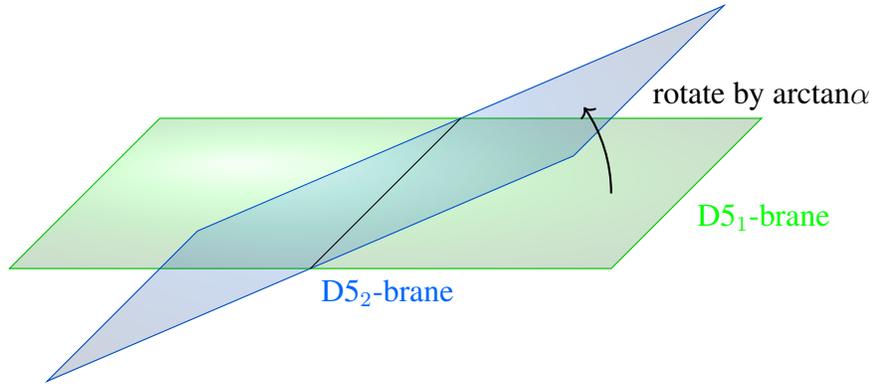
\end{center}

The fermion zero modes on this system are the stable normalizable solutions of the Dirac equations. The number of such solutions can be counted by the Callias index theorem\cite{Callias:1977kg}. The actual calculation can be found in \cite{Kanno:2021bze} for the mass term (\ref{mass-term}), and the result is 1 mod 2\footnote{
	There are three Weyl fermions with non-diagonal Yukawa coupling in our case. One can see that one linear combination of them does not arise in the Yukawa coupling, and two linear combinations of them form one Dirac fermion giving one Majorana fermion zero mode.
}. Therefore we have confirmed that this configuration has an odd number of fermion zero modes also in the field-theoretic argument. See also appendix \ref{sec:Callias-index} for Callias index theorem, and section \ref{sec:direct-index} for the explicit form of the zero modes in the $2+1$-dimensional case.

Compactifying this field theory along 5-direction, we get a point-like object with an odd number of Majorana fermion zero modes in the $4+1$-dimensional field theory on D5$_1$-brane. 
However, this field theory is not inconsistent because the object is infinitely heavy also in the field theoretical picture. Since in field theory there is no way to compactify 5-9 directions (which do not exist in the field-theoretic viewpoint!), we simply conclude that the field theory has a configuration with an odd number of fermion zero modes, but it is infinitely heavy and therefore non-dynamical.


\chapter{Conclusion}\label{chap:conclusion}


In this thesis, we studied the fermion zero modes on solitons and its relation to the non-perturbative anomalies. We further discussed the fermion zero modes on branes in string and M theories. Let us review the contents below.

\bigskip

In chapter \ref{chap:anom-QFT}, we gave a brief review of anomalies in quantum field theory. Anomaly polynomial gives a unified understanding of perturbative anomalies. We introduced the Witten anomaly as an example of non-perturbative anomalies. We also introduced the bordism group, which gives the mathematical classification of non-perturbative anomalies.

In chapter \ref{chap:anom-string}, we reviewed how the anomalies in string theories cancel. Thanks to the Green-Schwarz mechanism and modularity of internal CFT, string theories have no perturbative anomalies. On the other hand, non-perturbative anomalies in string theories are not fully understood. It is an open question what internal CFTs give consistent string vacua that are free from non-perturbative anomalies.

In chapter \ref{chap:MFZ}, we introduced the concept of solitons in quantum field theory, and reviewed the effects of fermion zero modes on solitons. The meaning of having an odd number of Majorana fermion zero modes crucially depends on the spacetime dimension: In $1+1$ and $2+1$ dimensions, it is a sign of the existence of non-abelian statistics. In $3+1$ and higher dimensions, it signifies an inconsistency.

In chapter \ref{chap:consistency}, we gave a general argument that having an odd number of fermion zero modes on a point-like soliton signifies an inconsistency in $3+1$ or higher dimension, under some mild assumptions. The inconsistency is likely to be related to the $\bZ_2$-valued non-perturbative anomalies, such as the Witten anomaly.

In chapter \ref{chap:branes-wrap}, we investigated the fermion zero modes on wrapped branes in string and M theories. We showed that a single (coincident) M5-brane(s) always has an even number of fermion zero modes in terms of the R-symmetry on the worldvolume theory, and in terms of the geometric calculation. 

In chapter \ref{chap:branes-intersect}, we calculated the number of fermion zero modes on the intersecting branes. We could construct models with an odd number of Majorana fermion zero modes for $\#\text{ND}=8$, but not for other $\#\text{ND}$. When there is an odd number of fermion zero modes on a (coincident) brane(s), compactifying it to be point-like is not allowed. Finally, we gave a discussion that the gauge invariance of the worldvolume theories on coinciding branes always requires an even number of Majorana fermion zero modes, if the branes are compactified to be point-like.

\bigskip

Our original questions were
\begin{itemize}
	\item What kind of configurations of string theories are free from non-perturbative anomalies? What are the consistency conditions of string theory compactifications?
	\item Can we realize anyons in string theory?
\end{itemize}

For the first question, we proposed a new criterion; Having an odd number of Majorana fermion zero modes on a point-like brane is not allowed. We found an example of inconsistent configuration, D0-D8-O4$^+$ which is not known before to the author's knowledge. 
On the other hand, previously-known consistency conditions of string theories prevent us from constructing other examples. 
Furthermore, we found another interpretation of having an odd number of Majorana fermion zero modes in terms of the gauge invariance on the branes. 
We can conclude that string theory already prohibits an odd number of Majorana fermion zero modes on a point-like brane.

For the second question, from the discussion in section \ref{sec:interpret-brane-intersect}, we conclude that anyons of our interest cannot be realized by the wrapped branes and the intersecting branes. The realization of anyons in string theories should be more complicated, if possible.

Having that said, we investigated a limited class of vacua of string theories. One possible future direction is to study other vacua of string theories and see the fermion zero modes on solitons. 
The realization of an odd number of fermion zero modes is either a sign of the existence of anyons or a sign of inconsistency, depending on the spacetime dimensions. Both of the possibilities are worth investigating further.


\chapter*{Acknowledgment}


First and foremost, I am grateful to my supervisor Taizan Watari. As a supervisor and a collaborator, he spent a lot of time giving me lectures, guiding my research work, and doing paperwork for me. My first research project started with his advice to search the anomalies in string theory. I originally did not expect that I would continue to work on this subject, but things were going around, and it became one of the key concepts in my final work in the Ph.D. course.

\bigskip

I also would like to give special thanks to Yuji Tachikawa, who was my virtual supervisor during my work on this thesis. He has also been one of indispensable collaborators of mine throughout my Ph.D. course. I remember that his comments on my master's thesis urged me to collaborate with him, directing me to where I am now. I have no doubt that my activities in my Ph.D. course become fruitful thanks to his kind support and discussions.
\bigskip

I thank all of my colleagues at the University of Tokyo and Kavli IPMU. I enjoyed chats, discussions and activities with them, even after choosing different paths in life.  In particular, I would like to thank Yuichi Enoki as one of my collaborators. I also thank everyone who discussed with me and gave comments on my work. I cannot list all of them here.

\bigskip

I also appreciate the financial support from the JSPS Fellowship for Young Scientists, the IGPEES program and the WPI Initiative, and all the support from the offices at the University of Tokyo and Kavli IPMU. Without them I could not afford any travel expenses or items including my PC I am using to write this sentence right now.

\bigskip

Finally, let me take this opportunity to express my gratitude to my family, my wife, and all of my private friends. They have been giving me so much emotional support, and bringing color into my life.

\appendix

\renewcommand{\appendixname}{Appendix}

\chapter{Open String Spectrum}\label{app:open-spectrum}

In this chapter, we explain how to count the massless degrees of freedom on open strings ending on branes in type IIA or IIB theory.

The setups are as follows. There are two D-branes (or $N+M$ coincident D-branes) stretching on the flat $9+1$-dimensional spacetime. We assume that the two branes coincide along some directions, and intersect vertically in other directions. 

There are two possible boundary conditions for oscillators $X^{\mu}$ of an open string ending on branes: Dirichlet and Neumann boundary conditions. If boundary conditions of an oscillator $X^{\mu}$ on both ends are Dirichlet, the direction is called a DD direction. ND and NN directions are defined in the same way. The number of each direction is denoted as $\#\text{DD},\#\text{ND},\#\text{NN}$. They obviously satisfy $\#\text{DD}+\#\text{ND}+\#\text{NN}=10$ and $\#\text{NN}\geq1$.

The configuration of the D-branes is specified by $\#$NN, $\#$ND, $\#$DD. We calculate the massless (and tachyonic) states on an open string stretching between them. We assume that $\#$ND is even.

In section \ref{sec:bosonic-spectrum}, we explain the bosonic excitations. In sections \ref{sec:NS-spectrum} and \ref{sec:R-spectrum}, we explain the fermionic excitations in the NS/R sectors. In section \ref{sec:orientifold-spectrum}, we discuss the effects of orientifolds. In section \ref{sec:example-spectrum}, we show an example of the calculation. See the textbook \cite{Polchinski:1998rr} for more details for example.

\section{Bosonic Excitation}\label{sec:bosonic-spectrum}

We use $\tau,\sigma$ as coordinates of the worldsheet of an open string. $\tau$ is the time direction and $\sigma$ is the space direction. Our coordinate is such that the ends of a string are at $\sigma=0,\pi$. There are 10 bosonic fields $X^{\mu}$ on the worldsheet theory, corresponding to the spacetime direction.

The Neumann boundary condition is
\begin{align}
    \partial_{\sigma}X^{\mu}=0
\end{align}
at the boundary. The Dirichlet condition is
\begin{align}
    \partial_{\tau}X^{\mu}=0
\end{align}
at the boundary.

\subsubsection*{NN Direction}

This can also be used for open strings ending on the same branes on both sides. The general solution of the equation of motion and the boundary condition is
\begin{align}
    X^{\mu}(\tau,\sigma)=x^{\mu}+p^{\mu}\tau+c\sum_{n\neq 0}\frac{\alpha^{\mu}_n}{n}{\rm cos}(n\sigma)e^{in\tau}
\end{align}
where $c$ is a constant. There are two weight 0 operators $x^{\mu},p^{\mu}$ and an infinite number of operators $\alpha^{\mu}_n$ with weight $n$.

\subsubsection*{DD Direction}

The general solution is
\begin{align}
    X^{\mu}=x^{\mu}+\delta^{\mu} \sigma+c\sum_{n\neq 0}\frac{\alpha^{\mu}_n}{n}{\rm sin}(n\sigma)e^{in\tau}.
\end{align}

\subsubsection*{DN and ND Direction}

The general solution is
\begin{align}
    X^{\mu}=c\sum_{r\in 1/2+\bZ}\frac{\alpha^{\mu}_r}{r}{\rm sin/cos}(r\sigma)e^{ir\tau}
\end{align}
where sin/cos corresponds to DN and ND. The difference from above is that the weights of operators $\alpha^{\mu}_r$ become $1/2+\bZ$.

\section{NS Sector}\label{sec:NS-spectrum}

There are fermions $\psi^{\mu}$ in the worldsheet theory. In the NS sector they transform in the opposite way as the bosons under the transformation $\sigma\rightarrow\sigma+2\pi$. Fermions expand as
\begin{align}
    {\rm NN} &: \psi^{\mu}=c\sum_{r\in 1/2+\bZ}\frac{\psi^{\mu}_r}{r}{\rm cos}(r\sigma)e^{ir\tau},\\
    {\rm DD} &: \psi^{\mu}=c\sum_{r\in 1/2+\bZ}\frac{\psi^{\mu}_r}{r}{\rm sin}(r\sigma)e^{ir\tau},\\
    {\rm DN/ND} &: \psi^{\mu}=\psi^{\mu}_0+c\sum_{n\neq 0}\frac{\psi^{\mu}_n}{n}{\rm sin/cos}(n\sigma)e^{in\tau}
\end{align}
in the NS sector.

The ground states $\ket{0,k}_{\rm NS}$ of the worldsheet are defined so that they are annihilated by oscillators with weights greater than 0. The zero-point energy of the ground states depends on the number of zero modes, and that is
\begin{align}
    (8-\# {\rm ND})(-\frac{1}{24}-\frac{1}{48})+\# {\rm ND}(\frac{1}{48}+\frac{1}{24})=-\frac{1}{2}+\frac{\# {\rm ND}}{8}.
\end{align}

Because $\ket{0,k}_{\rm NS}$ has $e^{i\pi F}=-1$, and we must choose states with $e^{i\pi F}=1$ as physical states by the GSO projection, the bosonic states are made by multiplying any number of bosonic oscillators and an odd number of fermionic oscillators. Massless bosons arise only when $\#$ND$=0,4$. 

When $\#$ND=0, which is the situation where the open string ends on the same brane on both sides, the massless states are
\begin{align}
    \psi^{\mu}_{1/2}\ket{0,k}_{NS}
\end{align}
which correspond to gauge bosons and collective coordinates on the branes.

When $\#$ND=4, the massless bosonic state is a Weyl representation of $SO(4)_{\rm ND}$. There are two real degrees of freedom.

\section{R Sector}\label{sec:R-spectrum}

In the R sector fermions transform in the same way as the bosons under $\sigma\rightarrow\sigma+2\pi$.
The expansion of fermions in the R sector is the same as bosons:
\begin{align}
    {\rm NN} &: \psi^{\mu}=\psi^{\mu}_0+c\sum_{n\neq 0}\frac{\psi^{\mu}_n}{n}{\rm cos}(n\sigma)e^{in\tau},\\
    {\rm DD} &: \psi^{\mu}=\psi^{\mu}_0+c\sum_{n\neq 0}\frac{\psi^{\mu}_n}{n}{\rm sin}(n\sigma)e^{in\tau},\\
    {\rm DN/ND} &: \psi^{\mu}=c\sum_{r\in 1/2+\bZ}\frac{\psi^{\mu}_r}{r}{\rm sin/cos}(r\sigma)e^{ir\tau}.
\end{align}

The zero-point energy of the ground states is zero because of the cancellation between bosons and fermions. Fermion zero modes (NN and DD) act on the ground states. Because the time-direction is NN, it gives massless fermions on the brane.

The physical conditions are the GSO projection and the so-called $G_0$-condition, which restricts $s_0=\pm1/2$ for left and right-movers\footnote{
    We assume that we can take a light-cone gauge. Therefore, the analysis cannot be directly applied to the cases such as $\#$NN=1. Nevertheless the result will be applicable to such cases, because $\#$NN changes under T-duality without changing physics.
}.

As a result, the fermionic degrees of freedom on the open string is\footnote{
    This number is the number of Majorana-Weyl fermions in a light-cone $1+1$ dimension. In other words, the number of $0+1$-dimensional Majorana fermions.
}
\begin{align}
    2^{(\#{\rm DD}+\#{\rm NN})/2}/2=2^{4-\#{\rm ND}/2}.
\end{align}
When the open string stretches between different D-branes, each orientation has this degree of freedom.

\section{Orientifolds}\label{sec:orientifold-spectrum}

When there exists an orientifold $r$-plane O$r$, the spectrum of an open string ending on D-branes is identified under
\begin{align}
    \label{ori-id}
    u(\tau,\sigma)\ket{0,k,ij}_{R/NS}\rightarrow \gamma^{-1} R(u\ket{0,k,ji}_{R/NS})\gamma
\end{align}
where $u$ is some operator made from oscillators and $i,j$ denotes each end of the open string. $\gamma$ acts on Chan-Paton indices, and
\begin{align}
    \gamma=1
\end{align}
or
\begin{align}
    \gamma=i\begin{pmatrix}
        0&1\\
        -1&0\\
    \end{pmatrix}=\sigma_2
\end{align}
acting on $i,j$.

$R$ is the action of the reflection of the spacetime. Let us denote the directions where both the O-plane and the D-brane stretch as $X^{\mu}$, only the O-plane stretches as $X^i$, only the D-brane stretches as $X^p$ and none of them stretches as $X^{\alpha}$. Then $R$  sends $X^{\mu,i}\rightarrow X^{\mu,i}$, and $X^{p,\alpha}\rightarrow -X^{p,\alpha}$. The extension of this action to fermionic ground states should respect worldsheet supersymmetry. It acts as
\begin{align}
    R=-{\rm exp}[i\pi\sum \mathbf{s}_{i,p}]
\end{align}
on fermionic (R-sector) ground states.

One can check that the massless RR closed string states are sent to themselves if $m=4n$ and sent to minus of themselves if $m=4n+2$ by the orientifold action, where $m$ is the difference between the rank of the RR anti-symmetric tensor field and the dimension $r+1$ of O$r$-plane. The D$p$-branes have RR charge of $C_{p+1}$, so the D$p$-branes also should be sent to themselves if $p-r=4n$ and minus of themselves if $p-r=4n+2$. What is the "minus" of the D-brane? It can be interpreted as the anti-brane, which is almost the same object but the sign of the RR charge flips. 
To summarize, if there exists an O$r$-plane, D$(r+4n)$-branes are sent to themselves and D$(r+4n+2)$-branes are sent to their anti-branes. 
For example, let us consider the type I superstring. The spacetime is filled with 32 D9-branes and an O9-plane. D1, D5, and D9-branes could exist alone. However D(-1), D3, and D7-branes cannot, because they alone are not invariant under the action of the O9-plane. The brane-anti-brane pair could exist because it is invariant. Actually some of them are stable\footnote{
    When two of such pairs exist, they will decay to nothing. Therefore the pair is said to have $\bZ_2$-charge.
}. The classification of possible charges of branes is done by K-theory \cite{Witten:1998cd}.

We will focus on the case where $N$ coincident D$p$-branes are on an O$r$-plane. Suppose the difference $p-r$ is divisible by 4 so that D-branes (not with anti-branes) alone can exist. The original spectrum of the coincident D-branes is a $U(N)$ gauge theory. The gauge fields come from the massless states $A_{\mu}^{ij}=\psi^{\mu}_{1/2}\ket{0,k,ij}_{NS}$. With the orientifold-plane, the gauge fields are identified under
\begin{align}
    A^{ij}_{\mu}\rightarrow-\gamma^{-1}A^{ji}_{\mu}\gamma.
\end{align}
When $\gamma=1$, the Lie algebra is restricted to the anti-symmetric matrices. The resulting theory is an $SO(N)$ gauge theory. When $\gamma=\sigma_2$, the resulting theory is an $Sp(N/2)$ gauge theory. Obviously $N$ has to be even in this case.

When there exist multiple (not coincident) D-branes, open strings can stretch between them. The spectrum on such open string is also identified under the orientifold projection (\ref{ori-id}), which relates the two orientations of the open string and as a result halves the degrees of freedom.

Finally, we comment on the two types of orientifolds, O$r^{\pm}$. As we discussed there are two types of the orientifold action, $\gamma=1$ and $\gamma=\sigma_2$. We define so that O$r^-$-plane acts on D$r$-brane with $\gamma=1$, and O$r^+$-plane with $\gamma=\sigma_2$. 

The actions of O$r^-$-planes on D$(r\pm4)$-branes is determined to be of $\gamma=\sigma_2$, O$r^+$-planes on D$(p\pm4)$ of $\gamma=1$, and so on. The reason is as follows.
Let us see the action of O$r^+$-plane on D$(r-4)$-brane. We consider coincident O$r^+$-plane and D$r$-brane, and a D$(r-4)$-brane contained in the O$r^+$-plane. An open string can stretch between the D$r$ and the D$(r-4)$-branes. 
Let there be two of such open string. When they approach and almost coincide, they can be interpreted as a single open string ending on D$r$-brane on both sides, and also ending on D$(r-4)$-brane on both sides. Therefore, the action of O$r^+$ is on D$r$, and also on D$(r-4)$, and they should match. Recall that the massless open string spectrum on the NS sector with $\#\text{ND}=4$ is constructed by $\psi_0^{a+}\ket{0,k}_{NS}$. The action of O$r^+$ on the state is 1 on D$r$ and $i$ on D$(r-4)$. Because we consider two of them, we get the relation
\begin{align}
	(\gamma_p^T)^{-1}\gamma_p=i^2(\gamma_{p-4}^T)^{-1}\gamma_{p-4},
\end{align}
which determines the action of O$r^+$ on D$(r-4)$.

\section{Example}\label{sec:example-spectrum}

We will show the example of the systematic calculations of the spectrum on open strings on D-branes. As an example we focus on the D2-O6$^+$ system, because this situation is one of our main interests and complicated enough to choose as an example. The setup is as follows. There is an O6$^+$-plane stretching along 0-6 directions, and on it, $N$ coincident D2-branes are along 0-2 directions. We are interested in the worldvolume theory of the D2-branes.

The bosonic massless spectrum is made from the states $\psi^{\mu}_{-1/2}\ket{0,k,ij}_{NS}$ because the zero-point energy is $-1/2$. The orientifold projection acts on them as
\begin{align}
	\psi^{\mu}_{-1/2}\ket{0,k,ij}_{NS}\rightarrow-\psi^{\mu}_{-1/2}\ket{0,k,ji}_{NS}
\end{align}
for $\mu=$0-2,7-9, and
\begin{align}
	\psi^{\mu}_{-1/2}\ket{0,k,ij}_{NS}\rightarrow\psi^{\mu}_{-1/2}\ket{0,k,ji}_{NS}
\end{align}
for $\mu=$3-6. Therefore, the bosonic spectrum is an $SO(N)$ gauge boson $A$, three real scalars $\Phi^a$ in the adjoint representation of SO(N), and four real scalars $\phi^p$ in the symmetric representation of $SO(N)$.

The fermionic massless spectrum is made from the states $\ket{\mathbf{s},k,ij}_R$. The GSO projection and $G_0$ condition restrict the physical states so that
\begin{align}\label{GSO-G0}
	s_{01}=+1/2,\,\,\, {\rm exp}(i\pi\sum_{i=23}^{89}\mathbf{s}_i)=1.
\end{align}
The orientifold projection acts on them as
\begin{align}
	\ket{\mathbf{s},k,ij}_R\rightarrow\ket{R(\mathbf{s}),k,ji}_R
\end{align}
where
\begin{align}
	R=-{\rm exp}(i\pi\sum_{i=34,56}\mathbf{s}_i).
\end{align}
Taking the basis $\mathbf{s}=(s_{01},s_{34},s_{56},s_{78},s_{92})$, $R$ sends $(s_{34}=s_{56})$ to themselves, and $(s_{34}=-s_{56})$ to minus of themselves. Combined with the conditions (\ref{GSO-G0}), the fermionic spectrum is two Dirac fermions $\Psi^b$ in the adjoint representation of $SO(N)$, and two Dirac fermions $\psi^q$ in the symmetric representation of $SO(N)$\footnote{
	For each of $(s_{34}=\pm s_{56})$, $(s_{01}=+1/2,s_{92})$ form a $2+1$-dimensional Dirac fermion. $s_{78}$ is determined uniquely by the GSO projection.
}.

Let us see the symmetry of this theory. Besides the Lorentz symmetry and supersymmetry, this theory has $SO(4)\times SO(3)\simeq SU(2)_R\times SU(2)_L\times SU(2)_N$ symmetry corresponding to the spatial rotation of 3-6 and 7-9 directions. Given that this theory should have eight supercharges\footnote{
	Eight is the number of real supercharges. Originally we have 32 supercharges in $9+1$ dimension, and each of the existence of D2-branes and O$6^+$-plane halves the number of supercharges.
}, $SU(2)_R$ can be interpreted as the R-symmetry of the theory\footnote{
	Note that $R$ has double-meaning, the ``right-hand-side'' in $SO(4)\simeq SU(2)\times SU(2)$ and ``R''-symmetry.
}. $(A,\Phi,\Psi)$ forms the vector multiplet of $2+1$-dimensional $\cN=2$ supersymmetry, whereas $(\phi,\psi)$ forms the hypermultiplet. Charges of each field are summarized in Table \ref{table-charges-open}. This theory has Yukawa coupling Tr($\overline{\Psi}\Gamma^p[\phi^p,\psi])+cc.$\footnote{
	The combination $\overline{\Psi}\Gamma^p\psi$ transforms like a vector representation under SO(4). Therefore this coupling is indeed invariant under symmetries.
} because of the supersymmetry\footnote{
	This coupling comes from the original $2+1$-dimensional $\cN=4$ worldvolume theory of D2-brane before taking the orientifold projection. $2+1$-dimensional $\cN=4$ gauge theory is so symmetric that the possible Yukawa coupling is unique up to isomorphisms. An easy way to see this is to start from $9+1$-dimensional $\cN=1$ super Yang-Mills, whose action is uniquely
	\begin{align*}
		-\frac{1}{4g^2}{\rm Tr}(F^2)-\frac{i}{2g^2}{\rm Tr}(\overline{\lambda}\Gamma^MD_M\lambda)
	\end{align*}
	where $\lambda$ is a $9+1$-dimensional Majorana-Weyl fermion and $M=0,\cdots,9$. The theory on D2-branes is uniquely the dimensional reduction of this action.
}, where $p=3,4,5,6$.
\vspace{3cm}
\begin{table}[h]
	\caption{Summary of charges of the theory on D2-branes with an O6$^+$-plane.}
	\label{table-charges-open}
    \centering
    \begin{tabular}{|c|c|c|c|c|}
        \hline
        &$SO(1,2)$&$SU(2)_R$&$SU(2)_L$&$SU(2)_N$ \\
        \hline \hline
        $A$&real vector&$\mathbf{1}$&$\mathbf{1}$&$\mathbf{1}$\\
        $\Phi$&real scalar&$\mathbf{1}$&$\mathbf{1}$&$\mathbf{3}_{\rm vec}$\\
        $\Psi$&Dirac fermion&$\mathbf{2}$&$\mathbf{1}$&$\mathbf{2}$\\
        $\phi$&real scalar&$\mathbf{2}$&$\mathbf{2}$&$\mathbf{1}$\\
        $\psi$&Dirac fermion&$\mathbf{1}$&$\mathbf{2}$&$\mathbf{2}$\\
        \hline
    \end{tabular}
\end{table}

Let us rewrite the fields in the case $N=2$. This theory is an $SO(2)=U(1)$ gauge theory. 
The $2\times2$ matrix-valued fields should be rewritten in the language of $U(1)$ theory where irreducible representations are one-dimensional. The adjoint fields are rewritten as
\begin{align}
	A=\begin{pmatrix}
		0&A^{new}\\
		-A^{new}&0
	\end{pmatrix}.
\end{align}
The kinetic term of $A^{new}$ becomes the ordinary $U(1)$ kinetic term up to normalization. 
$\Phi,\Psi$ are also decomposed similarly, giving $\Phi^{new},\Psi^{new}$. The decomposition of the fields in the symmetric representation $\phi,\psi$ is more complicated. Let us define
\begin{align}
	\phi=\begin{pmatrix}
		\phi_0+\phi_1&\phi_2\\
		\phi_2&\phi_0-\phi_1
	\end{pmatrix}.
\end{align}
A short calculation shows that Tr$(D_{\mu}\phi)^2=(\partial_{\mu}\phi_0)^2+|(\partial_{\mu}+2iA^{new}_{\mu})(\phi_2+i\phi_1)|^2$. Therefore, a boson $\phi$ in the symmetric representation of $SO(2)$ is decomposed into a real neutral boson $\phi_0$ and a complex boson $\phi^{new}_+:=\phi_2+ i\phi_1$ with $U(1)$-charge $2$. $\phi^{new}_-:=\phi_2- i\phi_1$ has $U(1)$-charge $-2$. A Dirac fermion $\psi$ in the symmetric representation also gives a neutral Dirac fermion $\psi_0^{new}$, a Dirac fermion $\psi_+^{new}$ with $U(1)$-charge $2$ and a Dirac fermion $\psi_-^{new}$ with $U(1)$-charge $-2$\footnote{
	Note that the 2 Dirac fermions $\psi_+$ and $\psi_-$ are independent of each other, whereas a complex boson $\phi_-$ is the complex conjugate of $\phi_+$. We denote $\overline{\psi_+},\overline{\psi_-}$ as the Dirac conjugates of $\psi_+,\psi_-$. We also note that $\overline{\psi_+}$ is not related to $\psi_-$.
}. We will omit the superscript $new$ for simplicity. The Yukawa coupling becomes $\overline{\Psi}\Gamma^p\phi^p_+\psi_-+\overline{\Psi}\Gamma^p\phi^p_-\psi_++cc.$

Concluding, the worldvolume theory of two D2-branes with an O6$^+$-plane is the $2+1$-dimensional $U(1)$ gauge theory with $\cN=2$ supersymmetry with the Lagrangian density
\begin{align}\notag
	\cL=&F^{\mu\nu}F_{\mu\nu}+(D_{\mu}\Phi)^2+(D_{\mu}\phi_0)^2+|D_{\mu}\phi_+|^2+2\overline{\Psi}\slashed{D}\Psi+\overline{\psi_0}\slashed{D}\psi_0\\
	&+\overline{\psi_-}\slashed{D}\psi_-+\overline{\psi_+}\slashed{D}\psi_++\overline{\Psi}\Gamma^p\phi^p_+\psi_--\overline{\Psi}\Gamma^p\phi^p_-\psi_++cc.
\end{align}
We have omitted the coefficients of each term. Each field has indices of $SU(2)_R\times SU(2)_L\times SU(2)_N$ and the sum should be taken accordingly.


\chapter{Fermions in Spacetime}\label{app:fermions}


This chapter explains the computation of fermion representations in Euclidean and Lorentzian signatures.
We will focus on the representations of the Clifford algebra
\begin{align}
    \{\Gamma^{\mu},\Gamma^{\nu}\}=\eta^{\mu\nu}
\end{align}
where $\eta^{\mu\nu}$ is the metric on the flat spacetime. Fermions are the sections of the bundle of some representation of the Clifford algebra.

In section \ref{sec:Euclidean-fermions}, we explain the representations in the Euclidean spacetime. In section \ref{sec:Lorentzian-fermions}, we explain the representations in the Minkowski spacetime.

\section{Euclidean}\label{sec:Euclidean-fermions}

The signature is $(d,0)$ and $\eta^{\mu\nu}=\delta^{\mu\nu}$. 

\subsection{Even Dimensions}

Suppose $d$ is even: $d=2k$. We define
\begin{align}
    \label{defGpm}
    \Gamma^{a\pm}:=\frac{1}{2}(\Gamma^{2a}\pm i\Gamma^{2a-1})
\end{align}
for $a=1,\cdots,k$\footnote{
	Sometimes we label them like $a=12,34,\cdots,(d-1)d$. 
}. They satisfy
\begin{align}
    \{\Gamma^{a+},\Gamma^{b-}\}&=\delta^{ab}\\
    \{\Gamma^{a+},\Gamma^{b+}\}&=\{\Gamma^{a-},\Gamma^{b-}\}=0.
\end{align}
Let $\zeta$ be a spinor annihilated by all $\Gamma^{a-}$:
\begin{align}
    \Gamma^{a-}\zeta=0, \ \ \ ^{\forall} a.
\end{align}
Then by acting $\Gamma^{a+}$ on $\zeta$, one can obtain $2^k$ dimensional complex representation. We call this representation the Dirac representation. It is convenient to set a basis $\mathbf{s}=(s_1,\cdots,s_k)$ to label them. $\zeta$ has the label $\mathbf{s}=(-1/2,\cdots,-1/2)$ and $\Gamma^a$ increases the label $s_a$ by 1. We will denote the basis as $\zeta^{\mathbf{s}}$. $s_a$ are the eigenvalues of the operators
\begin{align}
	S_a:=\Gamma^{a+}\Gamma^{a-}-\frac{1}{2}.
\end{align}

\subsubsection*{Weyl Spinor}

Define
\begin{align}
    \Gamma:=i^k\Gamma^1\cdots \Gamma^d.
\end{align}
$\Gamma$ satisfies
\begin{align}
    \Gamma^2=1,\ \ \ \{\Gamma,\Gamma^{\mu}\}=0.
\end{align}
The Dirac representation is decomposed by the eigenvalue of $\Gamma$ which is $\pm 1$. These representations are called the Weyl representations. The eigenvalues $\pm 1$ of Weyl representations are called chirality.
The Weyl representations are irreducible complex representations.

\subsubsection*{Majorana Spinor}
%


Defining
\begin{align}
    B_1:=\Gamma^1\Gamma^3\cdots \Gamma^{d-1}, \ \ \ B_2:=\Gamma B_1,
\end{align}
one can compute that
\begin{align}
    \label{MC1}
    B_1\Gamma^{\mu}B_1^{-1}&=(-1)^k\Gamma^{\mu *},\ \ B_2\Gamma^{\mu}B_2^{-1}=(-1)^{k-1}\Gamma^{\mu *},\\
    \label{MC2}
    B_1\Gamma B_1^{-1}&=B_2\Gamma B_2^{-1}=(-1)^k\Gamma^*=(-1)^k\Gamma.
\end{align}

Some spinors satisfy the Majorana condition
\begin{align}
    (\zeta^{\mathbf{s}})^*=B\zeta^{\mathbf{s}}
\end{align}
for $B=B_1$ or $B_2$. This condition is compatible with the algebra because $\Gamma^{\pm}$ is real. To satisfy this, we get a condition $\zeta^{\mathbf{s}}=B^*(\zeta^{\mathbf{s}})^*=B^*B\zeta^{\mathbf{s}}$, which means $B^*B=1$. By definition,
\begin{align}
    B_1^*B_1=(\Gamma^1\Gamma^3\cdots\Gamma^{d-1})^*\Gamma^1\Gamma^3\cdots\Gamma^{d-1}=(-1)^{k(k+1)/2},\ \ \ B_2^*B_2=(-1)^{k(k-1)/2}.
\end{align}
Therefore, to satisfy $B^*B=1$, we need $k=0,3$ mod $4$ for $B=B_1$ and $k=0,1$ mod $4$ for $B=B_2$. Therefore, Dirac representation can be decomposed by imposing the Majorana condition if $d=0,2,6$ mod $8$. These representations are called the Majorana representations. The Majorana condition halves the degree of freedom of a representation.

If $k$ is even, the Weyl condition is compatible with the Majorana condition. Therefore, the Majorana-Weyl representation is possible when $k=0$ mod $8$.

\subsubsection{Charge Conjugation}

The charge conjugation matrix $C$ has the property
\begin{align}
    C\Gamma^{\mu}C^{-1}=-\Gamma^{\mu T}.
\end{align}
The matrices $-\Gamma^{\mu T}$ also satisfy the Clifford algebra.

The operators $B_1,B_2$ are related to the charge conjugation matrix $C$ via the following relations
\begin{align}
    C=B_1\ \ {\rm for}\ \ k=1 \ \ {\rm mod}\ 2,\ \ \ C=B_2\ \ {\rm for}\ \ k=0\ \ {\rm mod}\ 2.
\end{align} 
$C$ satisfies $C^2=1$.

\subsubsection{Pseudo-Majorana Spinor}

Even if the Majorana representation is not possible when $d=6$ mod $8$, there are some ``real'' representations called pseudo-Majorana representations.

Consider two copies of the Dirac representation $\zeta_1^{\mathbf{s}},\zeta_2^{\mathbf{s}}$. Then imposing
\begin{align}
    (\zeta_1^{\mathbf{s}})^*=B\zeta_2^{\mathbf{s}},\ \ \ (\zeta_2^{\mathbf{s}})^*=B\zeta_1^{\mathbf{s}}
\end{align}
makes sense because $(B^*B)^2=1$. However, pseudo-Majorana representations are different from Majorana representations in that the pseudo-Majorana condition does not halve the dimension of the representation.

\subsection{Odd dimensions}

The representation of odd-dimensional ($2k+1$-dimensional) Clifford algebra can be realized by adding $\Gamma^{2k+1}=\Gamma$ to the $2k$-dimensional Clifford algebra. The complex dimensions of the Dirac representations are $2^k$. Since $\Gamma^{2k+1}=\Gamma$ is already in the algebra, there is no chirality and the Weyl representation.

\subsubsection*{Majorana Spinor}

The conditions (\ref{MC1}) and (\ref{MC2}) are compatible for only $B_1$. We cannot construct $B_2$ consistently. Therefore, Majorana conditions are possible when $k=0,3$ mod 4, namely $d=1,7$ mod 8.
Pseudo-Majorana spinors are possible as in the even-dimensional cases.

\subsection{Summary of spinors of $SO(d)$}

\begin{table}[h]
	\caption*{Summary of spinors of $SO(d)$}
    \centering
    \begin{tabular}{|c|c|c|c|c|}
        \hline
        d&Weyl&Majorana&Majorana-Weyl&minimum real-dim of rep. \\
        \hline \hline
        1&&yes&&1 (Majorana)\\
        2&yes&yes&&2 (Weyl or Majorana)\\
        3&&&&4 (Dirac)\\
        4&yes&&&4 (Weyl)\\
        5&&&&8 (Dirac)\\
        6&yes&yes&&8 (Weyl or Majorana)\\
        7&&yes&&8 (Majorana)\\
        8&yes&yes&yes&8 (Majorana-Weyl)\\
        9&&yes&&16 (Majorana)\\
        10&yes&yes&&32 (Weyl or Majorana)\\
        11&&&&64 (Dirac)\\
        12&yes&&&64 (Weyl)\\
        \hline
    \end{tabular}
\end{table}

\newpage

\section{Lorentzian}\label{sec:Lorentzian-fermions}

The signature is $(d-1,1)$ and $\eta^{\mu\nu}={\rm diag}(-1,1,\cdots,1)$. The representation analysis is almost the same as the Euclidean cases. 

\subsection{Even Dimensions}

In even dimensional cases $d=2k+2$, we define
\begin{align}
    \Gamma^{0\pm}&:=\frac{1}{2}(\Gamma^0\pm\Gamma^1),\\
    \Gamma^{a\pm}&:=\frac{1}{2}(\Gamma^{2a}\pm i\Gamma^{2a+1})
\end{align}
for $a=1,\cdots,k$.
The Dirac representations are complex $2^{k+1}$-dimensional.

We define the chirality matrix
\begin{align}
    \Gamma:=i^k\Gamma^0\Gamma^1\cdots\Gamma^{d-1}
\end{align}
satisfying the same relations as before. Dirac representations are reduced to the Weyl representations. They are complex irreducible representations.

The complex conjugation matrices $B_1, B_2$ are defined as before. The only difference is that $\Gamma^1$ is now self-conjugate. Then
\begin{align}
    B_1:=\Gamma^3\Gamma^5\cdots\Gamma^{d-1},\ \ \ B_2:=\Gamma B_1
\end{align}
satisfy
\begin{align}
    B_1\Gamma^{\mu}B_1^{-1}&=(-1)^k\Gamma^{\mu *},\ \ B_2\Gamma^{\mu}B_2^{-1}=(-1)^{k-1}\Gamma^{\mu *},\\
    B_1\Gamma B_1^{-1}&=B_2\Gamma B_2^{-1}=(-1)^k\Gamma^*=(-1)^k\Gamma
\end{align}
and
\begin{align}
    B_1^*B_1=(\Gamma^1\Gamma^3\cdots\Gamma^{d-1})^*\Gamma^1\Gamma^3\cdots\Gamma^{d-1}=(-1)^{k(k+1)/2},\ \ \ B_2^*B_2=(-1)^{k(k-1)/2}.
\end{align}
Therefore, the Majorana condition is possible when $k=0,3$ mod 4 for $B_1$ and $k=0,1$ mod 4 for $B_2$. In total, $d=0,2,4$ mod 8. Majorana and Weyl conditions are compatible when $k$ is even. So the Majorana-Weyl condition is possible when $d=2$ mod 8.

\subsection{Odd Dimensions}

When $d=2k+3$, the Clifford algebra is realized by adding $\Gamma^{2k+2}=\Gamma$ to the $2k+2$-dim Clifford algebra. The complex dimensions of the Dirac representations are $2^{k+1}$.

Only $B_1$ is possible for a complex conjugation matrix. Therefore Majorana representations are possible for $k=0,3$ mod 4, that is $d=1,3$ mod 8. The pseudo-Majorana condition is possible for any dimensions.

\subsection{Summary of spinors of $SO(d-1,1)$}

\begin{table}[h]
	\caption*{Summary of spinors $SO(d-1,1)$}
    \centering
    \begin{tabular}{|c|c|c|c|c|}
        \hline
        d&Weyl&Majorana&Majorana-Weyl&minimum real-dim of rep. \\
        \hline \hline
        1&&yes&&1 (Majorana)\\
        2&yes&yes&yes&1 (Majorana-Weyl)\\
        3&&yes&&2 (Majorana)\\
        4&yes&&&4 (Weyl)\\
        5&&&&8 (Dirac)\\
        6&yes&&&8 (Weyl)\\
        7&&&&16 (Dirac)\\
        8&yes&yes&&16 (Weyl or Majorana)\\
        9&&yes&&16 (Majorana)\\
        10&yes&yes&yes&16 (Majorana-Weyl)\\
        11&&yes&&32 (Majorana)\\
        12&yes&yes&&64 (Weyl or Majorana)\\
        \hline
    \end{tabular}
\end{table}


\chapter{Index Theorems}\label{app:index-theorem}


In this chapter, we enumerate various types of index theorems. In section \ref{sec:AS-index}, we explain the ordinary Atiyah-Singer index theorem for compact manifolds. In section \ref{sec:APS-index}, we explain the Atiyah-Patodi-Singer index theorem for manifolds with boundaries. In section \ref{sec:Callias-index}, we explain the Callias index theorem for manifolds with backgrounds. In section \ref{sec:mod2-index}, we describe the mod 2 index theorem. In section \ref{sec:direct-index}, we show a direct computation of fermion zero modes by solving the Dirac equation.

\section{Atiyah-Singer Index Theorem}\label{sec:AS-index}

Let $(M,g)$ be an $n$-dimensional spin manifold. Suppose there is a vector bundle $F\rightarrow M$ and its connection on $M$, and the action of Clifford bundle $Cl(M)$ on $F$ which commutes with the connection. Then $F$ is decomposed as $F=S(M)\otimes E$, where $S(M)$ is the spin bundle on $M$. Then the twisted Dirac operator is defined by
\begin{align}
    D_E:=\sum_{1\leq i\leq n}e_i\cdot\nabla_{e_i} : \Gamma(M,S(M)\otimes E)\rightarrow \Gamma(M,S(M)\otimes E).
\end{align}
$e_i$ is the local normal orthogonal frame of $TM$ and $\nabla$ is the covariant derivative of $F=S(M)\otimes E$. This definition is local, but $D_E$ is well-defined globally. $D_E$ is a first-order elliptic operator.

Suppose $n$ is even, $n=2k$. Then $D_E$ is decomposed as
\begin{align}D_E=\left(\begin{array}{cc}
    0&D_E^-\\
    D_E^+&0
\end{array}\right), \ \ D_E^{\pm}:\Gamma(M,S^{\pm}(M)\otimes E)\rightarrow\Gamma(M,S^{\mp}\otimes E).
\end{align}
The index theorem for $D_E$ is
\begin{align}
    \label{indthm}
    {\rm ind}(D_E):={\rm dim}({\rm ker}D_E^+)-{\rm dim}({\rm ker}D_E^-)=\int_M\hat{A}(TM)ch(E).
\end{align}

\subsection*{Example}

Let us compute the number of fermion zero modes on a Weyl fermion on 6-d manifold $\bR^{1,1}\times X\subset M_{11}$, where $X$ is a 4-manifold. We calculate the fermion zero modes for the fermion living in the pseudo-Majorana representation of $S(N)$, where $N$ is the normal bundle of $\bR^{1,1}\times X$. This setup corresponds to the worldvolume theory on an M5-brane wrapped around $X$.

Under the decomposition $SO(1,5)\rightarrow SO(1,1)\times SO(4)$, the $5+1$-dimensional twisted Dirac operator $D_{S(N)}^6$ is decomposed as
\begin{align}
    D_{S(N)}^6=D^2+D_{S(N)}^4.
\end{align}
$D^2$ is the ordinary $1+1$-dimensional Dirac operator and $D_{S(N)}^4:\Gamma(X,S(X)\otimes S(N))\rightarrow\Gamma(X,S(X)\otimes S(N))$ is a twisted Dirac operator on $X$.
Then the Atiyah-Singer index theorem tells us that
\begin{align}
    \label{indD4}
    {\rm ind}(D_{S(N)}^4)&=\int_X\hat{A}(TX)ch(S(N))\\
    &=\int_X(1-p_1(X)/24)(4+p_1(N)/2)=\int_X[p_1(N)/2-p_1(X)/6]\\
    &=\int_X\iota^*p_1(M_{11})/2-\int_Xp_1(X)/3.
\end{align}
This dimension is in the complex basis.

A Weyl fermion on a $5+1$-dimensional manifold decomposes as follows:
\begin{align}
    \mathbf{8}^+_{\rm 5+1Weyl}\rightarrow(\mathbf{1}^+_{\rm 1+1MW},\mathbf{4}^+_{\rm 4Weyl})\oplus(\mathbf{1}^+_{\rm 1+1MW},\mathbf{4}^-_{\rm 4Weyl}),
\end{align}
where $\mathbf{8}^+_{\rm 5+1Weyl}$ is a $5+1$-dimensional Weyl fermion with chirality $+$ which has real dimension 8, and so on.
Therefore the difference of the number of Majorana-Weyl fermions is equal to (\ref{indD4})\footnote{
    Because of the pseudo-Majorana-Weyl condition, the dimension of the Atiyah-Singer index theorem (\ref{indthm}) should be regarded as the real dimension. Or instead, we could start from a $5+1$-dimensional Dirac fermion and get the same number of complex fermions. The $5+1$-dimensional pseudo-Majorana condition halves the degree of freedom.
}. As a result, the number of MW fermions on the torus is
\begin{align}
    \int_X\iota^*p_1(M_{11})/2-\int_Xp_1(X)/3.
\end{align}
For a $4$-dimensional spin-manifold $X$, $\int_Xp_1(X)$ is known to be a multiple of $6$\footnote{
    This fact is derived from the Atiyah-Singer index theorem of the ordinary Dirac operator on $X$;
    \begin{align}
        {\rm ind}(D)=\int_X\hat{A}(TX)=-\frac{1}{24}\int_Xp_1(TX).
    \end{align}
    $\int_Xp_1(TX)$ is a multiple of $24$ because the left-hand side is an integer.
}. Therefore, the number of Majorana fermion zero modes is equal to $\int_X\iota^*p_1(M_{11})/2$ mod $2$, which matches well with the conclusion of the section \ref{sec:geom-branes-wrap}.

\section{Atiyah-Patodi-Singer Index Theorem}\label{sec:APS-index}

The Atiyah-Patodi-Singer (APS) index theorem \cite{Atiyah:1975jf} is an extension of the Atiyah-Singer index theorem for manifolds with boundaries. It is used to discuss the eta invariant for the anomaly calculation in section \ref{sec:modern-QFT}.

Let $M$ be an $(n+1)$-dimensional manifold with the $n$-dimensional boundary manifold $\partial M=N$. We define the Dirac operator $D_M:\Gamma(M,S(M)\otimes E)\rightarrow\Gamma(M,S(M)\otimes E)$ on $M$ and its restriction to the boundary $D_N:=D_M|_N$. Suppose $M$ is isomorphic to $N\times [0,1]$ near the boundary. We further assume the so-called APS boundary condition, which will be explained later. Then the index of $D_M$ is calculated by the Atiyah-Patodi-Singer index theorem
\begin{align}
	{\rm ind}D_M=\eta_N(0)+\int_M\hat{A}(R)Ch(F).
\end{align}
$\eta_N$ is defined to be the analytic continuation of
\begin{align}
	\eta_N^0(s)=\frac{1}{2}\sum_{\lambda:\text{eigenvalue of $D_N$}}sgn(\lambda)|\lambda|^{-s}.
\end{align}
When the boundary manifold $X$ is empty, the Atiyah-Patodi-Singer index theorem reduces to the Atiyah-Singer index theorem.

The APS boundary condition imposes a non-local condition
\begin{align}
	(D_N+|D_N|)\phi|_N=0
\end{align}
for a section $\phi$ of $S^+(M)$. The physical interpretation of this condition can be found in \cite{Fukaya:2017tsq,Kobayashi:2021jbn}.

\section{Callias Index Theorem}\label{sec:Callias-index}

Next we introduce the index theorem of Callias\cite{Callias:1977kg,Bott:1978bw}. This theorem is useful for calculating fermion zero modes on monopole backgrounds for example.

\subsection*{Theorem}

Let $L:S(\bR^n)\rightarrow S(\bR^n)$ be a first order differential operator on $\bR^n$, $n$ odd, of the form
\begin{align}
	L=i\Gamma^i\partial_i+i\Phi(x).
\end{align}
We assume that $\Phi(x)$ is a Hermitian matrix and approaches a homogeneous function of order 0 as $r\rightarrow\infty$.
Then the index theorem of Callias tells us that
\begin{align}
	{\rm ind}(L):={\rm dim(Ker}L)-{\rm dim(Ker}L^{\dagger})=\frac{1}{2((n-1)/2)!}\left(\frac{i}{8\pi}\right)^{\frac{n-1}{2}}\int_{S^{n-1}}{\rm tr}\left[U(x)(dU(x))^{n-1}\right]
\end{align}
where $U(x)=\Phi(x)/|\Phi(x)|$\footnote{
	We define $|\Phi(x)|$ so that
	\begin{align*}
		\Phi^{\dagger}\Phi=|\Phi(x)|^2.
	\end{align*}
} and $S^{n-1}$ is the sphere in $\bR^n$ of radius $\infty$. 

\subsection*{Example}

We will calculate the number of fermion zero modes on an 't Hooft-Polyakov monopole background in the $3+1$-dimensional SU(2) gauge theory by using this theorem. Recall that the Dirac equation is
\begin{align}\label{Dirac-eq}
	(\slashed{D}-\Phi(x))\psi=0,
\end{align}
where $\Phi$ is given by
\begin{align}
	\Phi(x)=a\frac{x^iT_i}{r}f(r).
\end{align}
Taking the basis $\Gamma^{\mu}=(i\sigma_2\otimes1,\sigma_3\otimes\sigma_1,\sigma_3\otimes\sigma_2,\sigma_3\otimes\sigma_3)$ and rewrite the fermion as $\psi=e^{iEt}(\psi_+,\psi_-)^T$, the Dirac equation (\ref{Dirac-eq}) becomes
\begin{align}
	\begin{pmatrix}
		0&L\\
		L^{\dagger}&0
	\end{pmatrix}\begin{pmatrix}
		\psi^+\\
		\psi^-
	\end{pmatrix}
	=E\begin{pmatrix}
		\psi^+\\
		\psi^-
	\end{pmatrix}
\end{align}
where
\begin{align}
	L=i\sigma_iD_i+i\Phi.
\end{align}

Redefining such as $\psi_+^{new}=e^{i\int A}\psi$, $D_i$ reduces to the ordinary derivative $\partial_i$. Because $A_i\rightarrow0$ as $r\rightarrow\infty$, this redefinition does not affect the result. The number of fermion zero modes is equal to dim(Ker$L$)$+$dim(Ker$L^{\dagger})$. Therefore, the number of fermion zero modes is equal to ind$(L)$ mod 2.

Let us compute the index for a fundamental fermion, $T_i=\sigma_i/2$. By the theorem we get
\begin{align}
	{\rm ind}(L)&=\frac{i}{16\pi}\int_{S^2}{\rm tr}(UdUdU)\\
	&=\frac{i}{16\pi}\int_{S^2}{\rm tr}\left[\frac{2x^iT_i}{r}d\left(\frac{2x^jT_j}{r}\right)\wedge d\left(\frac{2x^kT_k}{r}\right)\right]\\
	&=\frac{i}{16\pi}\int_{S^2}{\rm tr}(T_i[T_j,T_k]/2)\left[\frac{2x^i}{r}d\left(\frac{2x^j}{r}\right)\wedge d\left(\frac{2x^k}{r}\right)\right]\\
	&=\frac{i}{16\pi}\int_{S^2}i(1/2)\epsilon_{ijk}/2\left[\frac{8x^idx^jdx^k}{r^3}\right]\\
	&=-\frac{1}{8\pi}\cdot 6\cdot2\pi\cdot2/3=-1.
\end{align}
Here we used $[T_i,T_j]=i\epsilon_{ijk}T_k$ and tr$(T_iT_j)=\delta_{ij}/2$. Note that $|\Phi(x)|=a/2$ at $r\sim\infty$. This result is consistent with \cite{Harvey:1996ur} and the discussion in section \ref{sec:3+1-MFZ}.

In general, the index for a Dirac fermion in spin-$T$ representation of $SU(2)$ in the charge-$N$ 't Hooft-Polyakov monopole background is\cite{Callias:1977kg}
\begin{align}
	{\rm ind}(L)=-N[T(T+1)+A]
\end{align}
where $A=0$ $(1/4)$ for an integer (half-integer) $T$.

\section{Mod 2 Index Theorem}\label{sec:mod2-index}

Let $E,F$ be a real vector bundles over an $n$-dimensional compact manifold $X$, and $D:C^{\infty}(X;E)\rightarrow C^{\infty}(X,F)$ be an elliptic operator with real coefficients. $D$ is called skew-adjoint if it satisfies $D^{\dagger}=-D$. The ordinary index dim(ker$D$)$-$dim(ker$D^{\dagger}$) obviously vanishes for a skew-adjoint $D$. However, 
\begin{align}
	{\rm ind}_1D:={\rm dim(ker}D)\,\,\,\, {\rm mod}\, 2
\end{align} 
is a topological invariant and is called the mod 2 index of the skew-adjoint operator $D$. $D$ defines an element $[\sigma(D)]_1\in KO^{-1}(TX)$\footnote{
	The real K-theory $KO(X)$ on $X$ is defined to be the Grothendieck group of the finite-dimensional vector bundles. Roughly speaking the Grothendieck group is a group made from a monoid and the formal inverse of the monoid operation. The operation of $KO(X)$ is the Whitney sum. $KO^n(X)$ is defined so that $KO^0(X)=KO(X)$, $KO^n(X)=KO^{n+8}(X)$, and $KO^{-n}(X)=KO(S^nX)$. $S^nX$ is the $n$-th suspension of $X$. The suspension is defined to be $SX:=(X\times[0,1])/(X\times\{0\},X\times\{1\})$. For example $S^n(\text{point})=S^n$. 
} as follows. Let $D'=\text{cos}\pi x+D\text{sin}\pi x:C^{\infty}(X\times S^1;E)\rightarrow C^{\infty}(X\times S^1;F)$. Then the pullback of $D'$, $\sigma(D'):\pi_E^*E\rightarrow\pi_F^*F\in \Gamma(S^1\times TX,Hom(E,F))$, defines an element $[\sigma]\in KO(S^1\times TX)$ that is trivial at $x=0$. It can be regarded as an element $[\sigma]_1\in KO(S^1(TX))\simeq KO^{-1}(TX)$.

The mod 2 index theorem states \cite{Atiyah:1971rm} that the mod 2 index coincides with the image of $[\sigma]_1$ by the topological index
\begin{align}
	\text{t-ind}_1:KO^{-1}(TX)\rightarrow KO^{-1}({\rm point})=\bZ_2.
\end{align}
This is the restriction of the ordinary index of elliptic operators parametrized by $S^1$:
\begin{align}
	\text{t-ind}:KO(S^1\times TX)\rightarrow KO(S^1).
\end{align}

The topological index of $D$ is very hard to compute in general. We just give some comments on the mod 2 index of $D^5$ that appeared in the Witten anomaly. The space is $S^4\times\bR$ and the gauge field configuration is $A$ at $t\sim-\infty$ and $A^{g_4}$ at $t\sim\infty$. We set $A=0$. Because at $t\sim\pm\infty$ the gauge field is 0 up to gauge transformation, we can compactify the space to $S^5$. The topology $Prin_G(S^5)$ of $G$-bundle over $S^5$ is classified by $\pi_4(G)$. The mod 2 index t-ind$_1$ gives the morphism from $Prin_G(S^5)$ to $KO^{-1}(\text{point})$. For $G=Sp(N)$, where $\pi_4(G)=\bZ_2$, the mod 2 index is the isomorphism between them.

\section{Direct Computation of Zero Modes}\label{sec:direct-index}

There are three ways to compute the number of fermion zero modes on D-branes. One is explained in appendix \ref{app:open-spectrum}. Another is to use the index theorems. The third is to directly solve the Dirac equations. This section is devoted to showing a detailed computation to solve the Dirac equations following the method in \cite{Jackiw:1981ee}. We will see the matching of the results of the three methods. It gives another confirmation of our results discussed in chapter \ref{chap:branes-intersect}.

We will show how to solve the Dirac equation for the example D2-O6$^+$ system discussed in the section \ref{sec:example-spectrum},
\begin{align}\notag
	\cL=&F^{\mu\nu}F_{\mu\nu}+(D_{\mu}\Phi)^2+(D_{\mu}\phi_0)^2+|D_{\mu}\phi_+|^2+2\overline{\Psi}\slashed{D}\Psi+\overline{\psi_0}\slashed{D}\psi_0\\
	&+\overline{\psi_-}\slashed{D}\psi_-+\overline{\psi_+}\slashed{D}\psi_++\overline{\Psi}\Gamma\phi_+\psi_--\overline{\Psi}\Gamma\overline{\phi_+}\psi_++cc.
\end{align}
This theory has $U(1)_G$ gauge symmetry and $SU(2)_R\times SU(2)_L\times SU(2)_N$ symmetry. Charges of fields are summarized in Table \ref{table-charges-index}.

\begin{table}[h]
	\caption{Summary of charges of the theory on two D2-branes with an O6$^+$-plane. This table is essentially the same as \ref{table-charges-open}, but in the language of $U(1)$.}
	\label{table-charges-index}
    \centering
    \begin{tabular}{|c|c|c|c|c|c|}
        \hline
        &$SO(1,2)$&$SU(2)_R$&$SU(2)_L$&$SU(2)_N$&$U(1)_G$ \\
        \hline \hline
        $A$&real vector&$\mathbf{1}$&$\mathbf{1}$&$\mathbf{1}$&adj.\\
        $\Phi$&real scalar&$\mathbf{1}$&$\mathbf{1}$&$\mathbf{3}_{\rm vec}$&adj.\\
        $\Psi$&Dirac fermion&$\mathbf{2}$&$\mathbf{1}$&$\mathbf{2}$&adj.\\
        $\phi_0$&real scalar&$\mathbf{2}$&$\mathbf{2}$&$\mathbf{1}$&neutral\\
		$\phi_+$&complex scalar&$\mathbf{2}$&$\mathbf{2}$&$\mathbf{1}$&$+2$\\
        $\psi_0$&Dirac fermion&$\mathbf{1}$&$\mathbf{2}$&$\mathbf{2}$&neutral\\
		$\psi_+$&Dirac fermion&$\mathbf{1}$&$\mathbf{2}$&$\mathbf{2}$&$+2$\\
		$\psi_-$&Dirac fermion&$\mathbf{1}$&$\mathbf{2}$&$\mathbf{2}$&$-2$\\
        \hline
    \end{tabular}
\end{table}

Let us calculate the fermion zero modes on the configuration
\begin{align}
	\phi^3_+=\phi^3_-=\frac{xf(r)}{r},\,\,\, \phi^4_+=\phi^4_-=\frac{yf(r)}{r}
\end{align}
where $f(r)$ behaves like
\begin{align}
	f(r)\xlongrightarrow{r\rightarrow 0}f_0r,\,\,\, f(r)\xlongrightarrow{r\rightarrow\infty}f_{\infty}.
\end{align}
In the D-brane picture, this configuration corresponds to 
\begin{align}
	\phi^3=\begin{pmatrix}
		\frac{xf(r)}{r}&0\\
		0&0
	\end{pmatrix},
	\,\,\,
	\phi^4=\begin{pmatrix}
		\frac{yf(r)}{r}&0\\
		0&0
	\end{pmatrix}
\end{align}
up to the center of mass. Near the origin, $\phi^3\sim f_0x$ and $\phi^4\sim f_0y$. This means that we rotated one of the branes along 3,4-directions by the degree arctan$f_0$. The two D-branes intersect at the origin. On the other hand, far from the origin, $\phi^3\sim f_{\infty}{\rm cos}\theta$ and $\phi^4\sim f_{\infty}{\rm sin}\theta$. The two D-branes are separated by the distance $f_{\infty}$. Focusing on the regions near the origin, this example is a $2+1$-dimensional version of the model that appeared in section \ref{sec:field-brane-intersect}.

We can calculate the number of Majorana fermion zero modes by the Callias index theorem.
It can also be calculated by the counting method in the appendix \ref{app:open-spectrum}, by taking the limit $f_0\rightarrow\infty$. The answer is that there are four Majorana degrees of freedom, charged under $SO(5)_{5-9}$. Our goal is to understand these four zero modes by directly solving the Dirac equations.

The Dirac equations are
\begin{align}
	2\slashed{D}\Psi+(\Gamma^3\phi^3_++\Gamma^4\phi^4_+)\psi_--(\Gamma^3\phi^3_-+\Gamma^4\phi^4_-)\psi_+&=0,\\
	\slashed{D}\psi_+-(\Gamma^3\phi^3_++\Gamma^4\phi^4_+)\Psi&=0,\\
	\slashed{D}\psi_-+(\Gamma^3\phi^3_-+\Gamma^4\phi^4_-)\Psi&=0.
\end{align}
The fermion zero modes are the stable normalizable solutions of these equations. Because they are stable, we ignore the time dependence of the fields. The derivatives $\slashed{D}$ should be understood to include only space derivatives.
Now, redefining the fields as
\begin{align}
	\Psi&=2\Gamma^3\Psi_{old},\\
	\lambda&=\psi_++\psi_-,\\
	\psi&=e^{-2i\int A}\psi_--e^{2i\int A}\psi_+
\end{align}
and using $\phi^{3,4}_+=\phi^{3,4}_-$, we get
\begin{align}
	\slashed{\partial}\Psi-(\phi^3_++\Gamma^3\Gamma^4\phi^4_+)\psi&=0,\\
	\slashed{\partial}\psi+(\phi^3_++\Gamma^4\Gamma^3\phi^4_+)\Psi&=0,\\
	\slashed{D}\lambda&=0.
\end{align}
The general solution for $\lambda$ is a constant, which is not normalizable. Therefore we need to find the normalizable solutions for $\Psi$ and $\psi$.

Note that $\Gamma^3\Gamma^4=-\Gamma^4\Gamma^3=2iS_{34}$. Decomposing $\psi=(\psi_+,\psi_-)^T$ and $\Psi=(\Psi_+,\Psi_-)^T$ by the eigenvalue of $S_{34}$, the equations become
\begin{align}
	\slashed{\partial}\Psi_+-\frac{e^{i\theta}f(r)}{r}\psi_+&=0,\\
	\slashed{\partial}\psi_++\frac{e^{-i\theta}f(r)}{r}\Psi_+&=0,\\
	\slashed{\partial}\Psi_--\frac{e^{-i\theta}f(r)}{r}\psi_-&=0,\\
	\slashed{\partial}\psi_--\frac{e^{i\theta}f(r)}{r}\Psi_-&=0.
\end{align}
We focus on the two equations of $\Psi_+,\psi_+$. The analysis for $\Psi_-,\psi_-$ is essentially the same. Taking the basis $(\Gamma^1,\Gamma^2)=(\sigma_1,\sigma_2)$, and decomposing the fermions as $\Psi_+=(\Psi_1,\Psi_2)$, the two equations become
\begin{align}
	e^{i\theta}(\partial_r+\frac{i}{r}\partial_{\theta})\Psi_2-e^{i\theta}f(r)\psi_1&=0,\\
	e^{-i\theta}(\partial_r-\frac{i}{r}\partial_{\theta})\psi_1+e^{-i\theta}f(r)\Psi_2&=0,\\
	e^{-i\theta}(\partial_r-\frac{i}{r}\partial_{\theta})\Psi_1-e^{i\theta}f(r)\psi_2&=0,\\
	e^{i\theta}(\partial_r+\frac{i}{r}\partial_{\theta})\psi_2+e^{-i\theta}f(r)\Psi_1&=0.
\end{align}
We take an ansatz $\Psi_1=e^{im\theta}\Psi_1(r)$, $\Psi_2=e^{in\theta}\Psi_2(r)$. The $\theta$-dependence of $\psi_{1,2}$ must be $\psi_1=e^{in\theta}\psi_1(r)$, $\psi_2=e^{i(m-2)\theta}\psi_2(r)$. The equations become
\begin{align}\label{eq-final}
	(\partial_r-\frac{n}{r})\Psi_2-f(r)\psi_1&=0,\\
	(\partial_r+\frac{n}{r})\psi_1+f(r)\Psi_2&=0,\\
	(\partial_r+\frac{m}{r})\Psi_1-f(r)\psi_2&=0,\\
	(\partial_r-\frac{m-2}{r})\psi_2+f(r)\Psi_1&=0.
\end{align}
First we focus on the pair $(\Psi_2,\psi_1)$. There should be two independent general solutions for them. The asymptotic behavior at $r\sim\infty$ is $\sim e^{\pm\mu r}$. Only one of them is normalizable. The behavior at $r\sim0$ is
\begin{align}
	\Psi_2&\sim r^n, r^{-n+2},\\
	\psi_1&\sim r^{n+2},r^{-n}.
\end{align}
Given that only some linear combination of them is normalizable at infinity, all of them should behave well at the origin. The only possibility is $n=0$. Therefore there is one normalizable solution of the equations (\ref{eq-final}). Note that this solution is complex. In other words, this solution is charged under $SU(2)_N$. On the other hand, the same analysis gives the conditions $m\leq0,m-2\geq0$ for $(\Psi_1,\psi_2)$. Therefore we have only one complex solution from $(\Psi_+,\psi_+)$. It turns out that there is also one complex solution from $(\Psi_-,\psi_-)$. 

As a result, we have four real solutions as expected. Furthermore, we see that these solutions form $SO(2)_{56}$ doublets because $\psi_+$ and $\psi_-$ have different eigenvalues of $S_{56}$.

\bibliographystyle{ytamsalpha}
\baselineskip=.95\baselineskip
\bibliography{ref}

\end{document}